\documentclass[%
reprint,
amsmath,amssymb,
aps,
floatfix,
]{revtex4-1}
\usepackage[percent]{overpic}
\usepackage{braket}
\usepackage{float}
\usepackage{graphicx}
\usepackage{hyperref}
\hypersetup{
    colorlinks=true,
    linkcolor=blue,
    filecolor=magenta,
    citecolor=blue,
    urlcolor=cyan,
}

\newcommand{\HM}[2][blue]{\textcolor{#1}{ #2}}

\newcommand{\q}{\textup{$\hat{Q}$}}
\newcommand{\p}{\textup{$\hat{P}$}}

\begin{document}
\title{Microscopic model for a spatial multimode generation based on Multi-pump Four Wave Mixing in hot vapours}

\author{H. M. Florez}
\email{hans.m@ufabc.edu.br}

\affiliation{Centro de Ciências Naturais e Humanas, Universidade Federal do ABC-UFABC, Santo André 09210-580, Brazil.
}

\begin{abstract}
Multipartite entanglement is an important resource for quantum information processing. It has been shown that it is possible to employ alkali atoms to implement single device multipartite entanglement by using nonlinear processes with spatial modes. This work presents the first microscopic description of such multi-mode generation with two-pump four wave mixing (4WM) in dense atomic media. We implement an extension of a double $\Lambda$ model for a single pump 4WM in order to describe the multi-mode generation with a two-pump configuration. We propose a Floquet expansion to solve the multimode gain amplification and noise properties. The model describes the angle and the two-photon dependency of the multimode generation and the quantum correlations among the modes. We investigate the entanglement properties of the system, describing the main properties of previous experimental observations. Such a microscopic description can be used to predict the gain distribution of modes and  the quantum correlation within a typical range of experimental parameters.
\end{abstract}


\maketitle

\section{Introduction}
Quantum entanglement is an important resource for several protocols in quantum information, such as quantum teleportation~\cite{Zeilinger97,Kimble98},  quantum secret sharing~\cite{Hillery99,Gisin01,Pan05,Weedbrook13,Kunchi18}, and quantum networks~\cite{Kimble08,Gisin02,Grangier03}. The development of large quantum networks can exploit multipartite entanglement~\cite{Anirban2023,Marcello23}, which has made remarkable progress in experimental implementation~\cite{Kunchi03,Furusawa04,Hamel14,Lam15,Marcelo09,Marcelo18}.
There have been several demonstrations of discrete variables (DV) and continuous variables (CV), employing linear and non-linear optics. In the context of CV,  optical parametric oscillators (OPO) have been the main source used to create large ensembles of multimode entangled fields in time and frequency~\cite{Pfister2014,Roslund2014,Furusawa2013,Ulrik19,Jing23}. 
However, the distribution of quantum information carried by such modes is not suitable among different locations, as they are contained within two particular beams.

An alternative way has been recently shown in refs.\cite{Jing2018,Jing2019,Jing2020}, where the possibility of creating spatial multi-mode correlation using a four wave mixing process with a two-pump scheme has been demonstrated, exploiting the quantum correlation from a 4WM process as demonstrated in ref.\cite{PDLett07}. An interesting feature of the system, besides the spatial distribution, is the reconfigurability of the multimode entanglement among the generated modes. Such a property allows the production of n-partite  entanglement by employing pump shaping distributed among the spatial modes, i.e., producing 2-partite, 3-partite, 4-partite, etc., with the same system \cite{Jing2020}. They have also shown that the atomic system is a flexible platform to construct a reconfigurable multimode state with orbital angular momentum \cite{Jing2022}. 

The theoretical description of such a multi-mode process in hot vapours is hitherto based on an effective non-linear third-order hamiltonian~\cite{Jing2020,Jing2022} and a phenomenological beam-splitter gain relation~\cite{PDLett08}. 
However, the frequency bandwidth and tunability of the  four wave mixing process are not captured by such a model. The phenomenological  approach cannot map the range of parameters from the experiment, such as Rabi frequency and optical detuning with respect to the atomic line, and how these parameters can change the distribution of gain of the modes. Furthermore, the effective hamiltonian cannot predict the exact effect of extra modes in the case of interaction with multiple modes.

On the other hand, the microscopic model proposed by Glorieux et al. \cite{Quentin10} allows us to describe the amplification process in terms of the parameters of the atom-light interaction, such as Rabi frequency, optical detunings, and spontaneous emission rate. More recently, ref.\cite{Raul25} has shown that the double-$\Lambda$ model with quantum light fields can be used to provide a proper description of the two-mode gaussian quantum state produced by the amplification process with atoms. The same model can also be used to describe all optical control of the state through a six-wave mixing (6WM) interaction~\cite{Hans2025_1}.

In this work, we present a microscopic description of multi-mode generation with multiple pumps, which goes beyond the effective approach. Following the double-$\Lambda$ approach in ref.\cite{Raul25}, this work presents  a detailed theoretical description of multi-mode four-wave mixing (M4WM) based on Heisenberg-Langevin equations in a double lambda system within a thick medium interacting with several pumps. We employ the Floquet expansion, which allows us to extend the two mode system into a multimode system. With this method, we obtain a complete description of the multimode generation in the frequency domain, taking into account the main interaction parameters, such as optical detuning, the angle between the pump fields, and the pump Rabi frequencies. 
Our model closely describes the gain distribution  observed in ref.\cite{Jing2018}, as well as the quantum properties in the frequency domain, such as the multimode squeezing in ref.\cite{Jing2019} and the multipartite entanglement shown in ref.\cite{Jing2020}. This type of model also provides us with physical insight of how the modes propagate through the medium and explores different scenarios in which we have additional pumps. It can also be adapted to additional optical control~\cite{Hans2025_1} and an orbital angular momentum reconfigurable multimode state\cite{Jing2022}. 

The paper is organised as follows. In Section~\ref{sec:M4WM}, we introduce the double $\Lambda$ system used in the atomic levels to obtain the 4WM process and the Floquet expansion to achieve multimode amplification. 
In Section~\ref{sec:Gain}, we show the gain distribution of the multimode state as a function of the two-photon detuning for different angles. Section~\ref{sec:Prop} the propagation of such modes for different atom-light interaction parameters. Section~\ref{sec:Noise} presents the results on the noise properties and the multimode squeezing. Section ~\ref{sec:Entanglement} presents the results on multimode entanglement and how it is distributed among the different modes.  
Section~\ref{sec:Conclusions} presents our conclusions.

\section{Multimode dynamics from the Heisenberg-Langevin equations\label{sec:M4WM}}
Our theoretical analysis is based on the model described in ref.\cite{Raul25,Hans2025_1}, which offers a consistent structure that allows the state tomography of gaussian state in a 4WM in the frequency domain. In this section, we extend the 4WM process into a more general multimode structure generated by a double pump coupling. In order to make a proper extension of the double lambda model, we first introduce the 4WM process with a single pump, and then we present the extension to the multimode case.

\subsection{4WM process with a single pump \label{sec:4WM_sgl_pump}}

\begin{figure}[h!]
\centering
\includegraphics[width=86mm]{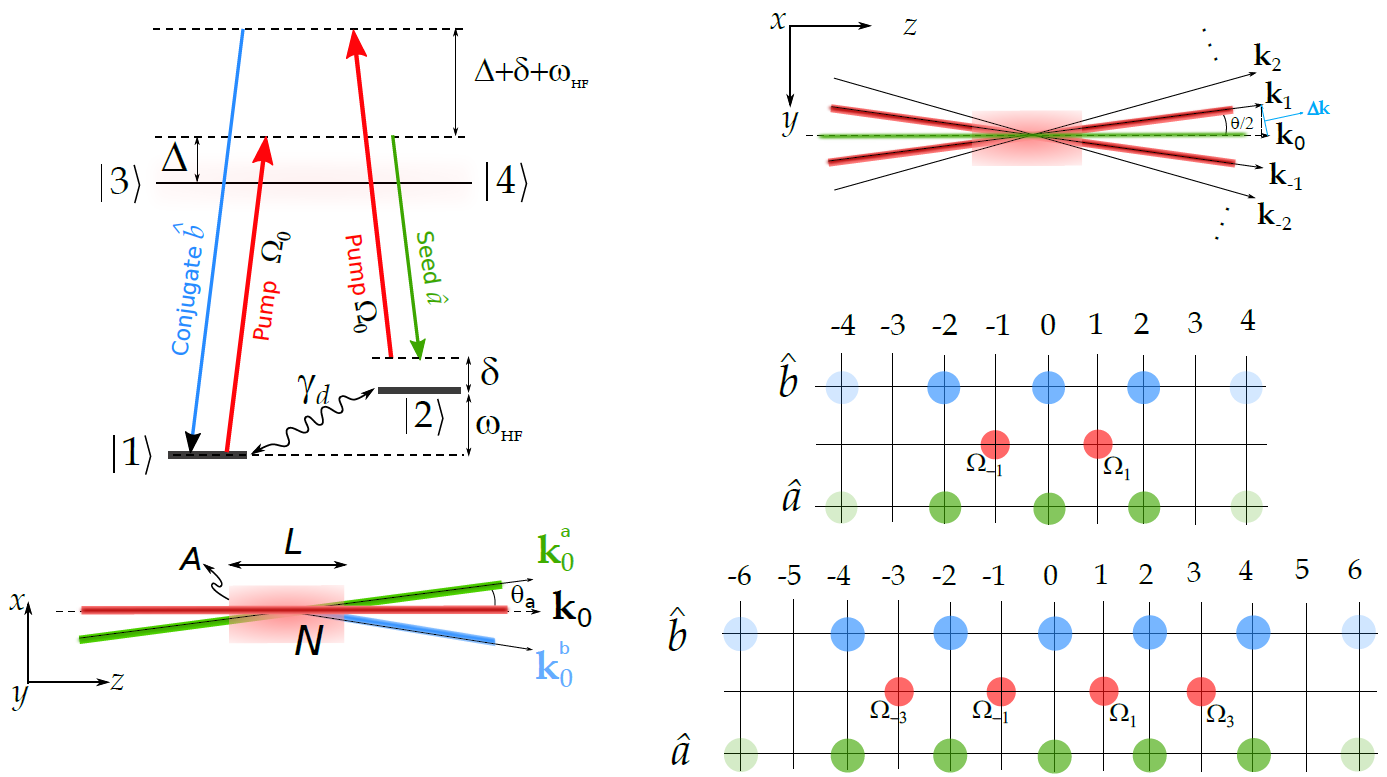}
\caption{(a) Energy level diagram with the double $\Lambda$ for a 4 level system. The hiperfine splitting is not resolve due to Doppler broadening. (b) Two-pump phase scheme. (c)
Multimode generation with a two-pump scheme and a generalization with four-pumps.}
\label{fig:Setup}
\end{figure}

Let us consider an ensemble of $N$ atoms within a cylinder with a cross section $A$ and a length $L$,  having a double $\Lambda$ type four level scheme of Fig.\ref{fig:Setup}(a) and (b), with two ground levels $|1\rangle$ and $|2\rangle$ and two excited levels $|3\rangle$ and $|4\rangle$. The two excited states are subjected to a spontaneous emission rate $\Gamma$, whereas the two ground states can decohere at a rate $\gamma$.  The single-pump 4WM case takes place along the $\mathbf{k}_0$ direction, with a probe beam slightly shifted from the $\mathbf{k}_0$ direction propagating towards $\mathbf{k}_0^a$ (see Fig.\ref{fig:Setup}(b)). The interaction of the pump and a probe field with frequencies $\omega_0$ and $\omega_0^a$
drives the lambda subsystem $|1\rangle\rightarrow |3\rangle\rightarrow |2\rangle$, whereas the interaction of the pump and the conjugate with frequencies $\omega_0$ and $\omega_0^b$ drives the second lambda subsystem $|2\rangle\rightarrow |4\rangle\rightarrow |1\rangle$.
The four wave  mixing (4WM) process annihilates two photons from the pump to generate an additional probe photon and a conjugate satisfying the phase matching condition $2\mathbf{k}_0=\mathbf{k}_0^a+\mathbf{k}_0^b$ and energy conservation $2\omega_0=\omega_0^a+\omega_0^b$. In general, we consider the interaction of the pump detuned by $\Delta$ with respect to the $|1\rangle\rightarrow |3\rangle$ transition, while the pump and the probe drive a Raman transition detuned by $\delta$. For the second $\Lambda$, the pump is detuned $\delta+\omega_{HF}$ and the conjugate by $\Delta+\omega_{HF}+\delta$.

The atomic hamiltonian is $\hat{H}_{a}=\int_0^L dz A \rho E_{i}\hat{\sigma}_{ii}(z)$ and the interaction hamiltonian for a single pump case and quantum fields $\hat{a}$ and $\hat{b}$ is 
\begin{align}
 \hat{H}_{int}=&\hbar \int_{0}^L\left(
 \Omega_0 e^{i(\omega_0t-\mathbf{k}_0\cdot\mathbf{r})}\hat{\sigma}_{31}+\Omega_0\hat{\sigma}_{42}e^{i(\omega_0t-\mathbf{k}_0\cdot\mathbf{r})}\right.\nonumber\\
 &\left.+g_0^a\hat{a}(z,t)\hat{\sigma}_{32}e^{i\omega_0^at}+g_0^b\hat{b}(z,t)\hat{\sigma}_{41}e^{i\omega_0^bt} + \text{h.c.}\right)\rho A dz,\label{eq:Hint_S_4WM}
 \end{align}
where $\Omega_0=\mathbf{d }\cdot\mathcal{E}_0/\hbar$ is the pump Rabi frequency and $g_0^a=p_{31}e_a$ and $g_0^a=p_{42}e_b$ are the atom-field coupling constants with $e_{a,b}=\sqrt{\hbar \omega_0^{a,b}/2\epsilon_0V}$ in which $V=A\ L$ is the quantization volume. The transition dipole moments are defined as $p_i$.

As described in ref.\cite{Quentin10} the evolution of the atomic operator is described by the Heisenberg-Langevin equations $\partial_t \hat{\sigma}_{ij}=-(i/\hbar)[\hat{\sigma}_{ij},\hat{H}_{int}] -\sum_{n,m}\Gamma_{nm}\hat{\sigma}_{nm}+\mathcal{\hat{F}}_{ij}$, where $\Gamma_{nm}$ describe spontaneous emission and decoherence rates, induced by the stochastic operators  $\mathcal{\hat{F}}_{ij}$ which satisfy $\langle\mathcal{\hat{F}}_{ij} \rangle=0$ and $\langle\mathcal{\hat{F}}_{ij}(z,t) \mathcal{\hat{F}}_{nm}(z',t')\rangle=\delta_{ij,nm}\delta(z-z')\delta(t-t')$.
Now, choosing $\Omega_i e^{i(\omega_i t-\mathbf{k}_i\cdot\mathbf{r})}$ with $i=1,2,3$ and 4, as our reference frame, we obtain the following dynamics in the Liouville space
 \begin{align}
\frac{d\mathbf{\hat{X}}(z,t)}{dt}=&\mathbf{M}^{(0)}
\mathbf{\hat{X}}(z,t)+\mathbf{G}_{x} \mathbf{\hat{A}}(z,t)+\mathcal{\hat{F}}(z,t),\label{eq:DynOnePumps_z}
\end{align}
where we have defined the atomic and stochastic vectors $\mathbf{\hat{X}}(z,t)=[\hat{\sigma}_{11},\hat{\sigma}_{12},\cdots,\hat{\sigma}_{44}]$ and $\hat{\mathcal{F}}(z,t)=[\hat{\mathcal{F}}_{11},\hat{\mathcal{F}}_{12},\cdots,\hat{\mathcal{F}}_{44}]$, respectively, while the light vector is defined as $\hat{\mathbf{A}}(z,t)=[\hat{a}(z,t),\hat{a}^\dagger(z,t), \hat{b}(z,t),\hat{b}^\dagger(z,t)]$. The matrix $\mathbf{G}_{x}$ depends on the atom-field coupling constants and the mean value of the atomic operators when atoms are only interacting with the pumps along $\mathbf{k}_0$.

The propagation of the probe  and conjugate fields $\hat{a}(z,t)$ and $\hat{b}(z,t)$ through the medium is given by 
the Heisenberg equation $(\partial_t + c\partial_z) \hat{O}=-(i/\hbar)[\hat{O},\hat{H}_{int}]$. 
such that the propagation of the fields is given by 
\begin{align}
(\partial_t + c\partial_z)\hat{\mathbf{A}}(z,t)=N\mathbf{T}\mathbf{\hat{X}}(z,t),\label{eq:Propg_OnePump}
\end{align}
where $N$ is the number of atoms, and the matrix $\mathbf{T}$ is a $4\times16$ of elements directly proportional to the atom-light coupling constants $g_0^a$ and $g_0^b$.

In order to obtain the gain profile, we consider the steady state of the atomic operators in eq.(\ref{eq:DynOnePumps_z})
 \begin{align}
\mathbf{\hat{X}}(z,t)=&-\mathbf{M}^{(0)-1}
\mathbf{G}_{x} \mathbf{\hat{A}}(z,t),
\label{eq:X_ss}
\end{align}
which leads to the propagation equation for the light field operator in the steady state condition ($\partial_t\hat{\mathbf{A}}(z,t)=0$)
\begin{align}
\frac{\partial \hat{\mathbf{A}}(z)}{\partial z}=\mathbf{R}\mathbf{\hat{A}}(z),\label{eq:dA_dz}
\end{align}
where $\mathbf{R}=N \mathbf{T}\mathbf{M}^{(0)-1}\mathbf{G}_x/c$ with the solution 
\begin{align}
\hat{\mathbf{A}}(z)=e^{\mathbf{R}\, z}\hat{\mathbf{A}}(0)\label{eq:A_z}.
\end{align}

This solution shows a simple relation between the light fields at the input $\hat{\mathbf{A}}(0)$ and output $\hat{\mathbf{A}}(z)$ of the interaction with the atomic medium. Given the initial state of the $\hat{a}$ and $\hat{b}$ modes, the propagation through the medium is represented by the propagator $\mathbf{J}=e^{\mathbf{R}\, z}$, reaching the new amplitudes at the output $\hat{\mathbf{A}}(z)$.  The atom-light parameters, such as intensities, optical detuning, and spontaneous emission rate, transform the matrix $\mathbf{R}$ in a way that modulates the input field and determines the tunability and bandwidth of the amplification process.

Now, the two-mode Gaussian state can be defined by the covariance matrix (CM), which contains quantum properties that fully describe the state of the light field modes. In order to obtain the CM,  we can employ the linearisation of the operator, i.e., the operator can be described as  $\hat{\mathbf{O}}=\langle \hat{\mathbf{O}}\rangle+\delta\hat{\mathbf{O}}$. In addition to the linearisation, we can describe the fluctuation in the frequency domain by applying the Fourier transform to the atomic and light operators, as shown in ref.\cite{Hans2025_1}, such as $\delta\hat{\mathbf{O}}(t)=\int d\omega \delta\hat{\mathbf{O}}(\omega)e^{-i\omega\, t}$, with $\omega$ as the analysis frequency, from which we obtain the solution 
\begin{align}
\delta\mathbf{\hat{A}}(z,\omega)&=\mathbf{J}(z,\omega)
\delta\mathbf{\hat{A}}(0,\omega)+\mathbf{J}(z,\omega) \mathcal{\hat{F}}_{in}(z,\omega),\label{eq:deltaA_omega}
\end{align}
where $\delta \mathbf{\hat{A}}(z,\omega)=\left[\delta a(z,\omega),\delta a^\dagger(z,\omega),\delta b(z,\omega), \delta b^\dagger(z,\omega)\right]$ and the propagator is defined as $\mathbf{J}(z,\omega)=\mathrm{exp}(\mathbf{R}(\omega)\, z)$ with $\mathbf{R}(\omega)=-(N/c)\mathbf{T}\, \tilde{\mathbf{M}}(\omega)\, \mathbf{G}_x +i(\omega/c) \mathbf{I}$, in which we have defined $\tilde{\mathbf{M}}(\omega)=[i\omega\mathbf{I}+\mathbf{M}]^{-1}$. The stochastic term is defined as $\mathcal{\hat{F}}_{in}(z,\omega)=(\hat{F}_{a}(z,\omega)$~$,\hat{F}_{a^\dagger}(z,\omega),\hat{F}_{b}(z,\omega),\hat{F}_{b^\dagger}(z,\omega))^T$ which is calculated from
\begin{align}
\mathcal{\hat{F}}_{in}(z,\omega)=&\int_0^z dz' e^{-\mathbf{R}(\omega)\ z'}\mathbf{R}_F(\omega) \mathcal{\hat{F}}(z',\omega),\label{eq:F_in}
\end{align}
with $\mathbf{R}_F(\omega)=(N/c)\mathbf{T}\, \tilde{\mathbf{M}}(\omega)$.

Therefore, this double-$\Lambda$ model shows that the interaction hamiltonian in eq(\ref{eq:Hint_S_4WM}) yields a straightforward solution for the quantum light fields from which we can obtain the gain spectrum from eq.(\ref{eq:A_z}) and the noise properties from eq(\ref{eq:deltaA_omega}) of the 4WM mixing process, as it is done in refs.\cite{Hans2025_1}.

\subsection{4WM process with two pump fields}
Now we are interested in the multiple pump case, and in particular, in the two-pump case where both pumps propagate along the directions $\mathbf{k}_{\pm1}=\mathbf{k}_0\pm \Delta \mathbf{k}$, as shown in Fig.\ref{fig:Setup}(c).
The angle given by $\Delta \mathbf{k}$ discretises the space of propagation modes. 
Each of the pump fields follows the classical description 
in which
$\hat{\mathbf{E}}_\pm(t)=\mathcal{E}_{\pm 1}e^{i(\omega_{\pm1} t-\mathbf{k}_{\pm1}\cdot\mathbf{r})} + c.c.$, such that these additional terms in the interaction Hamiltonian are described by 
$\hat{H}^{1}=\int_{0}^L dz\rho A \tilde{H}^{1}(z,t)$ with
\begin{align}
 \tilde{H}^{1}=&\hbar \left(
 \Omega_{-1} e^{i(\omega_0t-\mathbf{k}_{-1}\cdot\mathbf{r})}\hat{\sigma}_{31}+\Omega_1 e^{i(\omega_0t-\mathbf{k}_1\cdot\mathbf{r})}\hat{\sigma}_{31} + \text{h.c.}\right),
 \end{align}
where the projection of the wave vector along the propagation direction $\hat{z}$ is $\Delta k_z=[1-\cos(\theta_{\mathrm{eff}}/2)]\mathbf{k}_0$, as in Fig.\ref{fig:Setup}(c). The probe and conjugate modes are now referred to as channels such that $\hat{a}^{(n)}$ and $\hat{b}^{(n)}$ correspond to the modes of the probe and conjugate channels, respectively. An important aspect to consider is that the angle associated with the mismatch $\delta k$ needs to account for the angle between the pumps and the angle between the pump and the seed beam. Therefore, the effective angle is $\theta_{\mathrm{eff}}=((\theta/2)^2+\theta_a^2)^{1/2}$. Hence, 
by adding this Hamiltonian to the interaction Hamiltonian in eq.(\ref{eq:Hint_S_4WM}), i.e $\hat{H}_{int}+\hat{H}^{1}$,  the dynamics of the atomic operators can be written as
 \begin{align}
\frac{d\mathbf{\hat{X}}(z,t)}{dt}=&\left[\mathbf{M}^{(0)}+\mathbf{M}^{(1)}e^{i\Delta\mathbf{k}\cdot\mathbf{r}} +\mathbf{M}^{(-1)} e^{-i\Delta\mathbf{k}\cdot\mathbf{r}}\right]
\mathbf{\hat{X}}(z,t)\nonumber\\
&+ \mathbf{G}_{x} \mathbf{\hat{A}}(z,t)+\mathcal{\hat{F}}(z,t),\label{eq:DynTwoPumps_z}
\end{align}
where $\mathbf{M}^{(0)}$ describes  the single-pump case in the $\mathbf{k}_0$ direction, containing the main parameters of the interaction, whilst $\mathbf{M}^{(\pm 1)}$ represents the coupling with the two pumps which depend on $\Omega_{\pm 1}$.

This equation shows that in the case of $\Omega_{\pm 1}=0$ we obtain the single-pump dynamics along the $\mathbf{k}_0$ as in refs.~\cite{Quentin10,Raul25}, where the steady state of the operators is simply $\mathbf{\hat{X}}(z,t)=\mathbf{M}^{(0)-1}\mathbf{G}_x\mathbf{\hat{A}}(z,t)$. On the other hand, for the two-pump dynamics case when $\Omega_0=0$, the generators $\mathbf{M}_{\pm1}$  gain an extra phase $e^{\pm i \Delta\mathbf{k}\cdot\mathbf{r}}$ from which one can expand the solution.  Since the dynamics of the atomic operator is modified by the extra phase $e^{i\Delta k_z z}$, its steady state is not trivial as the case above. The extra phase affects the dynamics of the atomic operators as well as the propagation of the fields.


In order to solve the dynamics of this  two-pump interaction, we employ a Floquet expansion for the atomic and light operators, as 
 \begin{align}
\mathbf{\hat{X}}(z,t)=&\mathbf{\hat{X}}^{(0)}+\mathbf{\hat{X}}^{(1)}e^{i\Delta k_z z} +\mathbf{\hat{X}}^{(-1)} e^{-i\Delta k_z z}+\cdots\\
\mathbf{\hat{A}}(z,t)=&\mathbf{\hat{A}}^{(0)}+\mathbf{\hat{A}}^{(1)}e^{i\Delta k_z z} +\mathbf{\hat{A}}^{(-1)} e^{-i\Delta k_z z}+\cdots\\
\mathbf{G}_x(t)=&\mathbf{G}_x^{(0)}+\mathbf{G}_x^{(1)}e^{i\Delta k_z z} +\mathbf{G}_x^{(-1)} e^{-i\Delta k_z z}+\cdots\\
\mathcal{\hat{F}}(z,t)=&\mathcal{\hat{F}}^{(0)}+\mathcal{\hat{F}}^{(1)}e^{i\Delta k_z z} +\mathcal{\hat{F}}^{(-1)} e^{-i\Delta k_z z}+\cdots ,\label{eq:XFloqExp_z}
\end{align}
where the $n-$th element represents the atomic and light operators of each mode of propagation.
Assuming a finite expansion of modes up to $n=Q$, 
each atomic mode satisfies a recursive formula
\begin{align}
\frac{d\mathbf{\hat{X}}^{(n)}(z,t)}{dt}=&\mathbf{M}^{(0)}\mathbf{\hat{X}}^{(n)}+\mathbf{M}^{(1)}\mathbf{\hat{X}}^{(n-1)}+\mathbf{M}^{(-1)}\mathbf{\hat{X}}^{(n+1)}\nonumber\\
&+\sum_{i=-Q}^Q \mathbf{G}_x^{(n-i)} \mathbf{\hat{A}}^{(i)} + \mathcal{\hat{F}}^{(n)}, \label{eq:dXFloqExp}
\end{align}
as well as the  propagation of the light modes
\begin{align}
\frac{\partial \mathbf{\hat{A}}^{(n)}}{\partial t} +c\left[\frac{\partial }{\partial z}+i\ n \ \Delta k_z \right]\mathbf{\hat{A}}^{(n)}&=N\ \mathbf{T}\ 
\mathbf{X}^{(n)}(z,t).\label{eq:dAFloqExp}
\end{align}

The dynamics of each $\mathbf{\hat{X}}^{(n)}(z,t)$ term depend on itself and on the $\mathbf{\hat{X}}^{(n-1)}(z,t)$ and $\mathbf{\hat{X}}^{(n+1)}(z,t)$ terms. Moreover, the $n$-th term also has the contribution of all pairs of quantum fields $\mathbf{\hat{A}}^{(i)}$, which makes the algebraic solution non-trivial. Therefore, 
within a linear space of modes, we define the vectors $\mathbb{X}(z,t)$ and 
$\mathbb{A}(z,t)$ with elements $[\mathbb{X}(z,t)]_n=\mathbf{X}^{(n)}(z,t)$
and $[\mathbb{A}(z,t)]_n=\mathbf{A}^{(n)}(z,t)$. Hence, the dynamics of the atomic and light modes can be written in the space of modes as
\begin{align}
\frac{d\mathbb{\hat{X}}(z,t)}{dt}=&\mathbb{M}\mathbb{\hat{X}}+\mathbb{G}_x \mathbb{\hat{A}}(z,t) + \mathbb{\hat{F}} (z,t),\label{eq:XHarm}\\
\frac{\partial }{\partial t}\mathbb{\hat{A}}(z,t)& +c\left[\frac{\partial \mathbb{\hat{A}}(z,t)}{\partial z}+i\mathbb{N} \ \Delta k_z \mathbb{\hat{A}}(z,t)\right]=N\ \mathbb{T}\ \mathbb{X}(z,t),\label{eq:AHarm_z}
\end{align}
where we defined the matrix multi-mode generator with sub-matrices
\begin{align}
\mathbb{M}_{nm}=\begin{cases}
      \mathbf{M}^{(0)}, & \text{for}\ n=m \\
      \mathbf{M}^{(n-m)}, & \text{for}\ |n-m|=1\\
      0, &\text{otherwise},
    \end{cases}\label{eq:M_hiper},
\end{align}
the coupling matrices as $[\mathbb{G}_x]_{nm}=\mathbf{G}_x^{n-m}$ and $[\mathbb{T}]_{nm}=\delta_{nm}\mathbf{T}$, and $[\mathbb{N}]_{nm}=\delta_{nm}(-Q+n)\mathbf{I}_{4\times4}$ where $n,m$ correspond to the mode index.

These dynamical equations (\ref{eq:XHarm}) and (\ref{eq:AHarm_z}) for the atomic and light mode operators take an equivalent form as in the case of a single pump in eqs.(\ref{eq:DynOnePumps_z}) and (\ref{eq:Propg_OnePump}). Therefore, the algebraic solution is the same, and we can obtain the amplification gain and quadrature noise of the generated modes in the same way as in the single pump case.
The relevant aspect of this solution with respect to the effective approach is that matrices $\mathbb{M}$, $\mathbb{G}$, and $\mathbb{T}$ contain physical parameters such as frequency detuning, Rabi frequency, atom-field coupling constant, and spontaneous emission, which can be mapped directly to the experimental conditions. Furthermore, this approach can describe systems with a more complex structure, such as extra pumps, which are just elements of the diagonal in matrix $\mathbb{M}$. In the case of extra pairs of pumps, the coupling matrix can be expressed in the general form as
\begin{align}
\mathbb{M}_{nm}=\begin{cases}
      \mathbf{M}^{(0)}, & \text{for}\ n=m \\
      \mathbf{M}^{(n-m)}, & \text{for}\ |n-m|=1\\
      \vdots & \vdots \\
      \mathbf{M}^{(n-m)}, & \text{for}\ |n-m|=k\\
      0, &\text{otherwise},
    \end{cases}\label{eq:M_hiper},
\end{align}
where $k$ is the number of extra pairs of pumps. For instance, Fig.~\ref{fig:Setup}.(c) shows the case of $k=3$, in which the elements for the diagonal $|n-m|=2$ are zero and the elements of $|n-m|=3$ are $\mathbf{M}^{\pm3}$. 

\section{Gain of the optical modes\label{sec:Gain}}

In order to obtain the intensity gain distribution of the modes, we can first neglect the stochastic term, 
and consider the steady state condition $\partial_t \mathbb{X}=0$ and $\partial_t \mathbb{A}=0$, such that $
\mathbb{\hat{X}}^s=\mathbb{M}^{-1}\mathbb{G}_x \mathbb{\hat{A}}(z,t) $. Substituting in eq.(\ref{eq:AHarm_z}), the field at the output of the propagation is 
\begin{align}
\mathbb{\hat{A}}(z)&=e^{\mathbb{R}z}\mathbb{\hat{A}}(0), \label{eq:SolAHarm_z}
\end{align}
where we have defined 
$\mathbb{R}=(N\ \mathbb{T}\mathbb{M}^{-1}\mathbb{G}_x/c -i\mathbb{N} \ \Delta k_z)$ and the the $n-$th mode of this solution can be written as $[\mathbb{\hat{A}}(z)]_n=\hat{\mathbf{A}}^{(n)}=(\alpha^{(n)},\alpha^{*(n)},\beta^{(n)},\beta^{*(n)})$.



From this solution, we can obtain the gain matrix, which contains the gain of each mode. To do so, we define $\mathbb{C}(z)=\langle\mathbb{\hat{A}}(z),\mathbb{\hat{A}}(z)^T\rangle$ which in terms of the input mode fields is 
\begin{align}
\mathbb{C}(z)&=\mathbb{J}(z)~\langle\mathbb{\hat{A}}(0),\mathbb{\hat{A}}(0)^T\rangle~\mathbb{J}(z)^T,\label{eq:gain_modes_f}
\end{align}
as the gain matrix, where we have defined $\mathbb{J}(z)=e^{\mathbb{R}z}$, the input amplitude covariance matrix as $\mathbb{C}(0)=\langle\mathbb{\hat{A}}(0),\mathbb{\hat{A}}(0)^T\rangle$
and the submatrices  are given by $[\mathbb{C}(z)]_{nm}= \langle\mathbf{\hat{A}}^{(n)},\mathbf{\hat{A}}^{(m)T}\rangle$.
The amplitude covariance matrix at the input takes a particular form depending on the state of the fields before interaction with the atoms. In particular, the seed field in the probe channel can be considered as the mode $\mathbf{\hat{A}}^{(0)}$ in a coherent state, such that 
\begin{align}
\Braket{\mathbf{\hat{A}}^{(0)}(0),\mathbf{\hat{A}}^{(0)T}}&=\begin{bmatrix}
   \alpha^2 & |\alpha|^2 +1  & 0 & 0\\
    |\alpha|^2 & \alpha^{*2} & 0 & 0\\
    0 & 0 & 0 & 1\\
     0 & 0 & 0 & 0
\end{bmatrix},
\end{align}
and  $\langle \mathbf{\hat{A}}^{(n)}(0),\mathbf{\hat{A}}^{(m)T}(0)\rangle \approx0$ otherwise.
The intensity gain distribution  for each mode is  given by $G^{(n)}=\langle \mathbf{\hat{A}}^{(n)}(z),\mathbf{\hat{A}}^{(-n)T}(z)\rangle/|\alpha|^2$, which relates the mean of the number operator of each mode and compares it with the intensity of the seed beam. The calculation is done by choosing a total number of modes $n=2Q+1$ large enough to guarantee the convergence of the solution. To do so, we verify the number of modes $n=2Q+1$ that gives at least 1\% variation of the solution with respect to the $n=2(Q-1)+1$.

\begin{figure*}[t!]
\centering
\begin{overpic}[width=\textwidth]{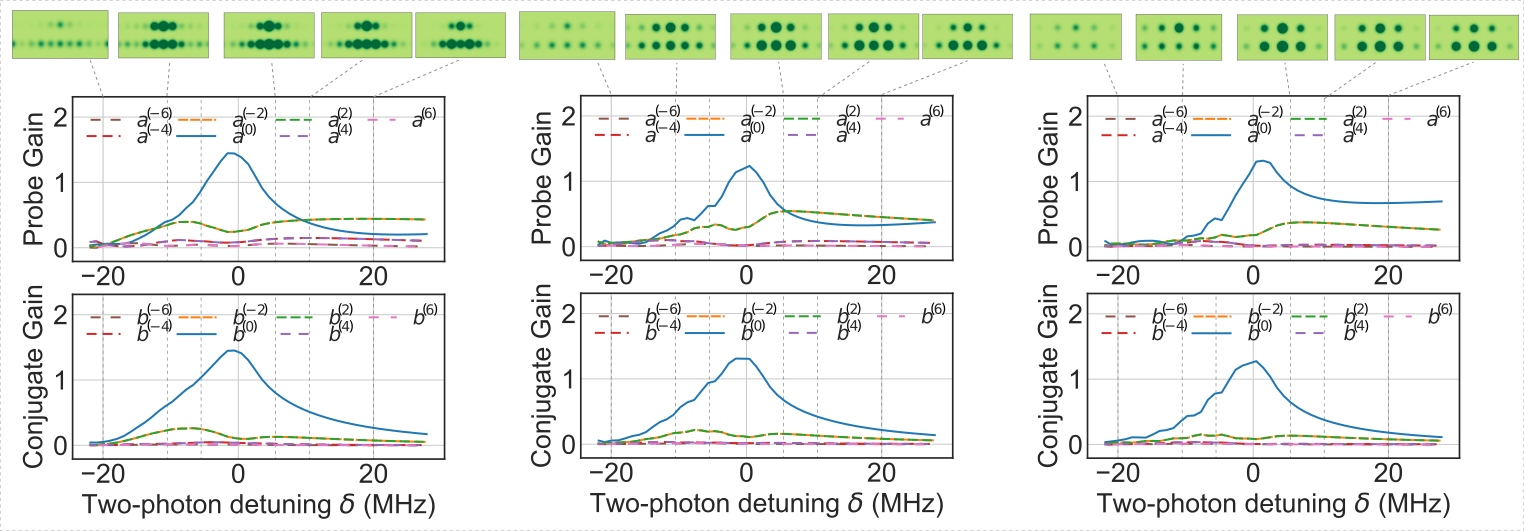}
\put(15,-2){(a)}
\put(48,-2){(b)}
\put(81,-2){(c)}
\end{overpic}
\vspace{0.1cm}
\caption{Gain amplification two-photon spectroscopy for different angles of the pump fields. Probe (upper plot) and conjugate (lower plot) gain distribution for (a) $\theta_{\mathrm{eff}}=3$~mrads, (b) $\theta_{\mathrm{eff}}=4.5$~mrads and (c) $\theta_{\mathrm{eff}}=6$~mrads. The parameters of the calculation are: $\Omega_0/2\pi=220$~MHz, $\Delta/2\pi=0.9$~GHz, $\Gamma/2\pi=5.7$~MHz,  $\gamma_d/2\pi=1$~MHz, $\omega_{\mathrm{HF}}/2\pi=3.035$~GHz and $g_a=g_b=0.28$~MHz. }
\label{fig:modes_gain_vs_d2}
\end{figure*}
\subsection{Gain spectroscopy}

Recently,  Shengshuai et al. in ref.\cite{Jing2019} reported an extensive characterisation of mode generation with a two-pump 4WM interaction using the D1 line of $^{85}$Rb. In that work, they showed that for large angles, the multimode process generates 6 modes; however, as the angle is reduced, the number of generated modes increases, exhibiting an asymmetric distribution with more modes in the probe channel than in the conjugate channel.

In our model, we consider the same transition at 795 nm, with ground states $|1\rangle$ and $|2\rangle$ as the states $F=2$ and $F=3$, respectively,  of $5~S_{1/2}$. The excited states $|3\rangle$ and $|4\rangle$ represent the states $F=2$ and $F=3$, respectively, of $5~P_{1/2}$, and we consider the spontaneous decay rate of levels $|3\rangle$ and $|4\rangle$ as $\Gamma_1/2\pi=5.7$~MHz. The  hyperfine splitting is $\omega_{HF}/2\pi=3.035$~GHz and  $\gamma_d/2\pi=$~1 MHz as the ground state decoherence.
The Rabi frequency of the pump beam can typically be used within the range of $0.1\leq\Omega_0/2\pi\leq 0.5$~GHz. In this case, we consider a pump beam of 100~mW  of power (typically lower than the single pump case) and 0.5~mm of waist size, so the Rabi frequency of the pump fields is $\Omega_0/2\pi=220$~MHz. The probe gain for a 0.3 mm beam waist is such that the coupling constants of the probe and conjugate fields are $g_a=g_b=0.28$~MHz. Since multimode generation involves fields with an increasing angle with respect to each other and the pump fields, the gain of the fields in eq.(\ref{eq:SolAHarm_z}) is integrated in the longitudinal direction ( see appendix~\ref{app:Doppler}) to account for the velocity distribution of the hot atomic ensemble. The integration does not alter the gain  distribution profile, it only smooths the curves for each mode.

Figure \ref{fig:modes_gain_vs_d2} shows the mode generation from the model, with the gain distribution as a function of the two-photon detuning for three different angles of the pump fields. 
The first case, Fig. \ref{fig:modes_gain_vs_d2}.(a), shows  the gain distribution for a small angle of $\sim$3.0 mrads. The spectrum indicates that for $\delta/2\pi>-10$~MHz, the process generates the modes in the probe channel $[\hat{a}^{(-6)},\hat{a}^{(-4)},\hat{a}^{(-2)},\hat{a}^{(0)},\hat{a}^{(2)},\hat{a}^{(4)},\hat{a}^{(6)}]$ and in the conjugate channel $[\hat{b}^{(-2)},\hat{b}^{(0)},\hat{b}^{(2)}]$. For $\delta/2\pi<-10$~MHz, more modes are populated, but with a lower gain.  Notice that the M4WM amplifies the even modes $\hat{a}^{(2n)}$ and $\hat{b}^{(2n)}$ since the pump fields act on the odd modes $\Omega_{\pm1}$ and the seed beam is in mode $\hat{a}^{(0)}$. An example of this kind of process is the amplification of $\hat{a}^{(2)}$. In this case, it can be produced by the annihilation of two photons of $\Omega_1$ and a photon from $\hat{b}^{(0)}$, or from one photon from each pump $\Omega_1$ and $\Omega_{-1}$ with a photon from $\hat{b}^{(-2)}$ that also produces a photon in $\hat{a}^{(2)}$. Similar combinations of 4WM processes with different $\Omega_{\pm1}$ occur for all even modes, whereas odd modes are not amplified because they do not satisfy the phase matching condition.

In order to visualise the transverse profile of amplified modes predicted by the model, in the upper part of Fig.~ \ref{fig:modes_gain_vs_d2}, we depict the transverse profile of the gain distribution as it would be seen by a CCD camera for different values of $\delta$. Notice that the gain distribution profile closely describes what is observed in ref.\cite{Jing2019}. For the case of a small angle, the number of modes populated in the probe channel is larger than that in the conjugate channel.

It is worth mentioning that the calculation is done with $Q=20$ such that the total number of modes is 41, showing a variation of 1\% with respect to the case of $Q=18$.  Therefore, we choose a large number of modes that guarantee the convergence of the solution for a lower number of excited modes, as plotted in Fig.~\ref{fig:modes_gain_vs_d2}. In appendix~\ref{app:Convergence}, we analyse the convergence for the angle of $\theta_{\mathrm{eff}}=3$~mrad, which is the one with the larger number of excited modes.

Now, as the angle increases, the model predicts a lower number of  generated modes. Figure~\ref{fig:modes_gain_vs_d2}.(b) shows  the gain distribution for an angle of 4.8 mrad, where 8 modes are mainly generated: 5 in the probe channel and 3 in the conjugate channel. However, we can now observe the weak presence of $\hat{a}^{(-4)}$ and $\hat{a}^{(4)}$ modes in the probe channel. This distribution mainly occurs for $\delta/2\pi>0$~MHz and near $\delta/2\pi\sim-10$~MHz.
Figure~\ref{fig:modes_gain_vs_d2}.(c), on the other hand, primarily shows 6 modes generated for a large angle of 6~mrads. In particular, for $\delta/2\pi>0$~MHz, the M4WM generates 6 intense modes, whereas for $\delta/2\pi<0$~MHz, the amplification is spread among the other modes.

It should be noted that the M4WM generates multiple modes in an asymmetric way in terms of the two-photon detuning, which is related to the dispersive gain profile of the 4WM; i.e., the generation of an intense beam occurs for $\delta>0$~MHz, whereas for $\delta<0$~MHz, the amplification is reduced as it approaches $\delta\sim-20$~MHz. Such a result is consistent with the observation in ref\cite{Jing2018,Jing2019}. 
Another aspect that the model predicts is that as the angle increases, a lower number of modes is amplified, in particular, 6 modes for an angle of $\theta_{\mathrm{eff}}=6$~mrad. To have a clearer perspective on this feature, Fig.~\ref{fig:modes_gain_vs_idx} shows the gain distribution for the three different angles at $\delta/2\pi \sim 5.5$~MHz, along with the respective transverse profile in the inset. Figures~~\ref{fig:modes_gain_vs_idx} (a-c) show the gain distribution for $\theta_{\mathrm{eff}}=3,4.8$ and 6~mrad, respectively. Figure (a) shows a gain distribution in the probe channel that spreads over modes up to $n=8$, going through a situation of 3 modes and 2 weak neighbours for $\theta_{\mathrm{eff}}=4.8$~mrads (b), until the situation of only  3 clear modes for each channel for $\theta_{\mathrm{eff}}=6$~mrads. Note that, while the gain distribution changes for the probe beam, the gain for the conjugate channel remains in 3 modes.

\begin{center}
\begin{figure}[t!]
\begin{overpic}[width=0.48\textwidth]{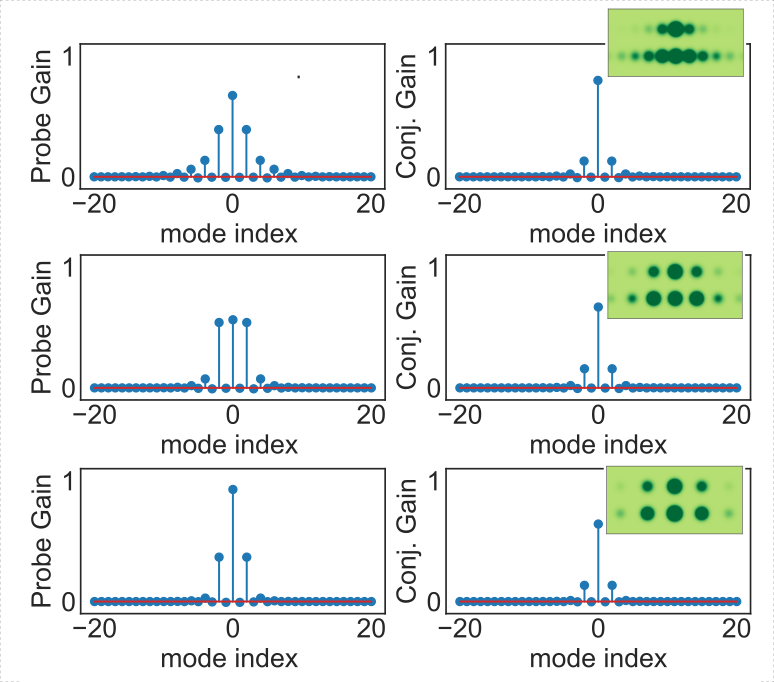}
\put(12,78){(a)}
\put(59,78){(b)}
\put(12,51){(c)}
\put(59,51){(d)}
\put(12,24){(e)}
\put(59,24){(f)}
\end{overpic}
\caption{Gain distribution for the probe and conjugate channel. (a) and (b) for $\theta_{\mathrm{eff}}=3$~mrads, (c) and (d) for $\theta_{\mathrm{eff}}=4.8$~mrads and (e) and (f) for $\theta_{\mathrm{eff}}=6$~mrads. The parameters of the calculation are: $\Omega_0/2\pi=220$~MHz, $\Delta/2\pi=0.9$~GHz, $\Gamma/2\pi=5.7$~MHz,  $\gamma_d/2\pi=1$~MHz, $\omega_{\mathrm{HF}}/2\pi=3.035$~GHz and $g_a=g_b=0.28$~MHz.}
\label{fig:modes_gain_vs_idx}
\end{figure}
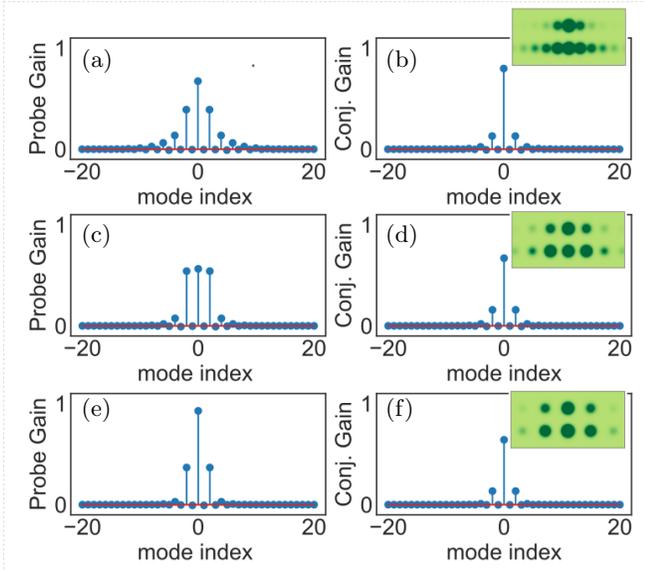
\end{center}

\section{Propagation of the fields through the medium\label{sec:Prop}}

\begin{figure}[h!]
\centering
\begin{overpic}[width=0.46\textwidth]{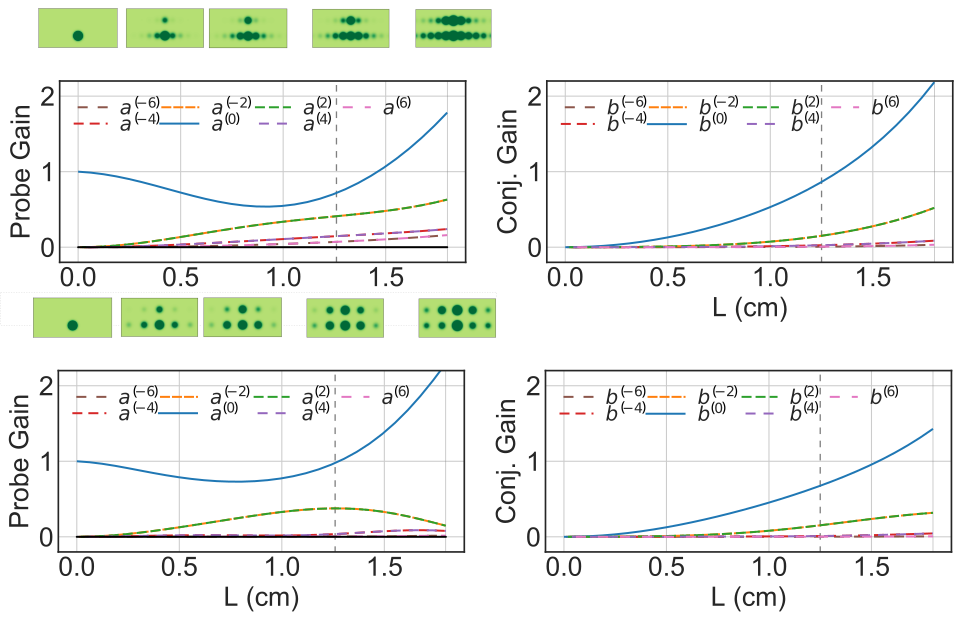}
\put(-3,57){(a)}
\put(-3,26){(b)}
\end{overpic}
\caption{Propagation of the modes for two different angles: (a) $\theta_{\mathrm{eff}}=3$~mrads and $\theta_{\mathrm{eff}}=6$~mrads. The parameters of the calculation are: $\Omega_0/2\pi=220$~MHz, $\Delta/2\pi=0.9$~GHz, $\Gamma/2\pi=5.7$~MHz,  $\gamma_d/2\pi=1$~MHz, $\omega_{\mathrm{HF}}/2\pi=3.035$~GHz and $g_a=g_b=0.28$~MHz. }
\label{fig:modes_propagation_d2_N5.5MHz}
\end{figure}

In addition to the characterisation of the gain distribution with respect to the angle and the two-photon detuning, the model offers physical insight from which we can analyse the multimode generation: the propagation through the medium. Figure~\ref{fig:modes_propagation_d2_N5.5MHz} shows the mode amplification process along the propagation through the medium for two different angles $\theta_{\mathrm{eff}}=3$~mrads and $\theta_{\mathrm{eff}}=6$~mrads, with $\delta/\pi = 5.5$ MHz. This  relevant aspect of the model allows us to observe how the seed beam excites the multimode process as it propagates through the medium and how the cascade multimode process occurs.

Figure~\ref{fig:modes_propagation_d2_N5.5MHz}(a) and (b) show that at the beginning of the cell for $L<0.5$~cm,  the probe beam (blue solid line) is attenuated, seeding the generation of the conjugate mode $\hat{b}_{0}$ (blue solid line in figures (c) and (d)).  This conjugate mode simultaneously seeds the first neighbour of the probe beam $\hat{a}^{(-2)},\hat{a}^{(0)},\hat{a}^{(2)}$ (orange and green dashed lines), which is clearly represented in the transverse profile (a.2) and (b.2). As the new modes start propagating within the range $0.5< L<1.0$~cm, they become seed fields for the modes $\hat{b}^{(-2)},\hat{b}^{(2)}$. This is possible through non-degenerate pump conversion, where one photon from each pump $\Omega^{\pm1}$ is employed to produce pairs of $(\hat{a}^{(\pm 2)},\hat{b}^{(\mp 2)})$. It is also important to notice that for a small angle $\theta=3$~mrads, the new modes $\hat{b}^{(-2)},\hat{b}^{(2)}$ simultaneously produce $\hat{a}^{(-4)},\hat{a}^{(4)}$, whereas for large angles $\theta=6$~mrads, those modes are not populated.

As the fields continuously propagate through the cell, they reach the typical cell size of $L=1.25$~cm (the vertical black dashed line)  used in this kind of experiment, where they attain the gain distribution discussed in Fig.\ref{fig:modes_gain_vs_d2}. That is the length in which small angles produce 10 modes with asymmetric gain for the probe and conjugate channels, whereas large angles produce 6 modes with a nearly symmetric distribution in the probe and conjugate channels. Now, instead, we consider a cell with a larger size, e.g., $L=1.8$~cm; some changes can be observed in the multimode amplification. In the case of small angles, the amplification across all modes is preserved and increases exponentially. Conversely, in the case of a large angle, the amplification of the modes $(\hat{a}^{(\pm 2)},\hat{b}^{(\mp 2)})$ in the probe channel is attenuated, while the amplification in the conjugate channel is reduced but still growing exponentially.

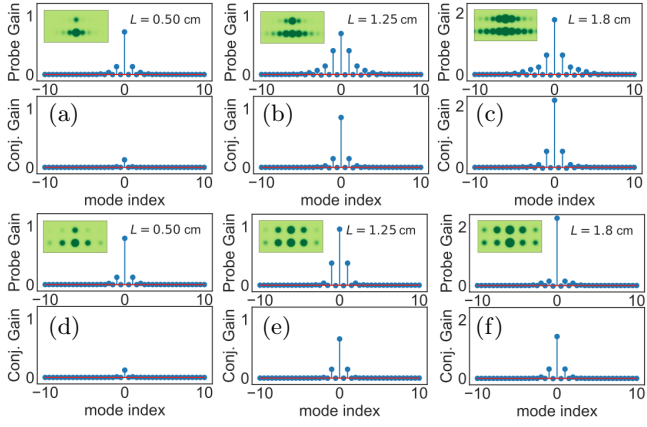
\begin{figure}[t!]
\centering
\begin{overpic}[width=0.48\textwidth]{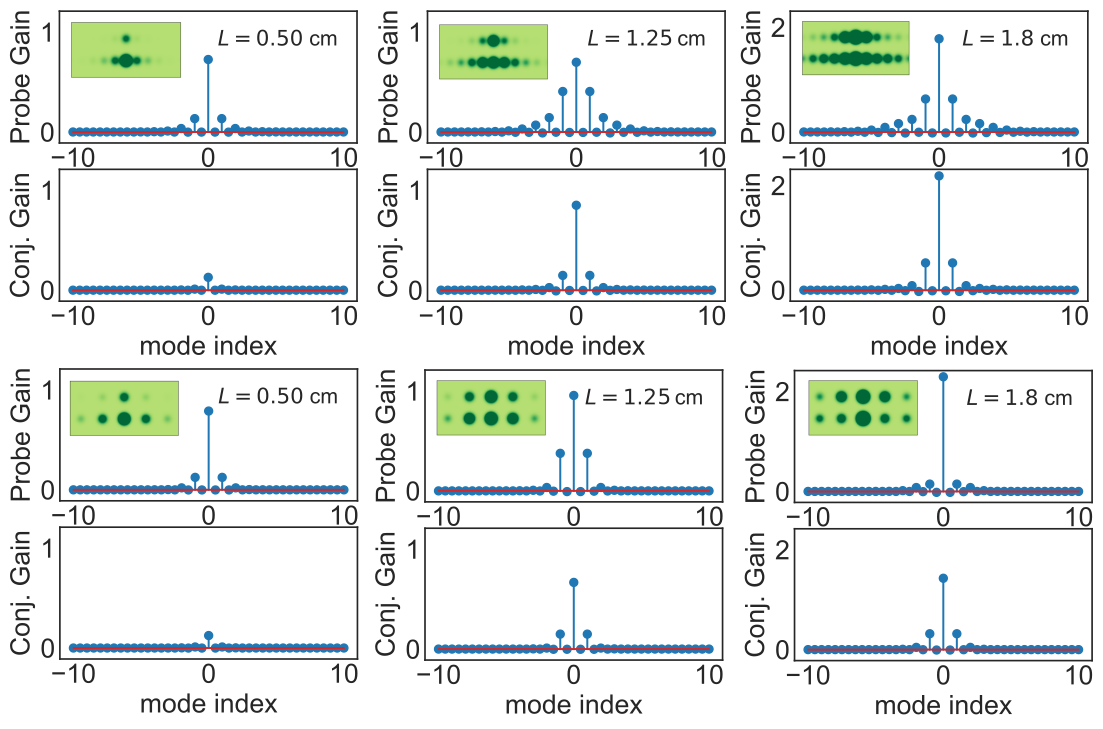}
\put(7,47){(a)}
\put(40,47){(b)}
\put(73,47){(c)}
\put(7,14){(d)}
\put(40,14){(e)}
\put(73,14){(f)}
\end{overpic}
\caption{Gain distribution for the probe and conjugate channel. (a) for $\theta_{\mathrm{eff}}=3$~mrads  and (b) for $\theta_{\mathrm{eff}}=6$~mrads. The parameters of the calculation are: $\Omega_0/2\pi=220$~MHz, $\Delta/2\pi=0.9$~GHz, $\Gamma/2\pi=5.7$~MHz,  $\gamma_d/2\pi=1$~MHz, $\omega_{\mathrm{HF}}/2\pi=3.035$~GHz and $g_a=g_b=0.28$~MHz.}
\label{fig:modes_idx_propagation_d2_N5.5MHz}
\end{figure}

In order to obtain a more quantitative perspective on the gain distribution, Fig.\ref{fig:modes_idx_propagation_d2_N5.5MHz} shows the gain distribution for 3 sections of the propagation for both angles.
Figure~\ref{fig:modes_idx_propagation_d2_N5.5MHz}~(a), (b) and (c) show the buildup of the multimode process for 3 different lengths, from  a $3\times1$ gain distribution of the probe and conjugate to a $7\time 3$ distribution. In contrast, for large angles, Figs.(d), (f), and (g) show the amplification for the same cell sizes, from a  $3\times1$ to a $1\times3$ distribution.



\begin{figure}[t!]
\centering
\begin{overpic}[width=0.46\textwidth]{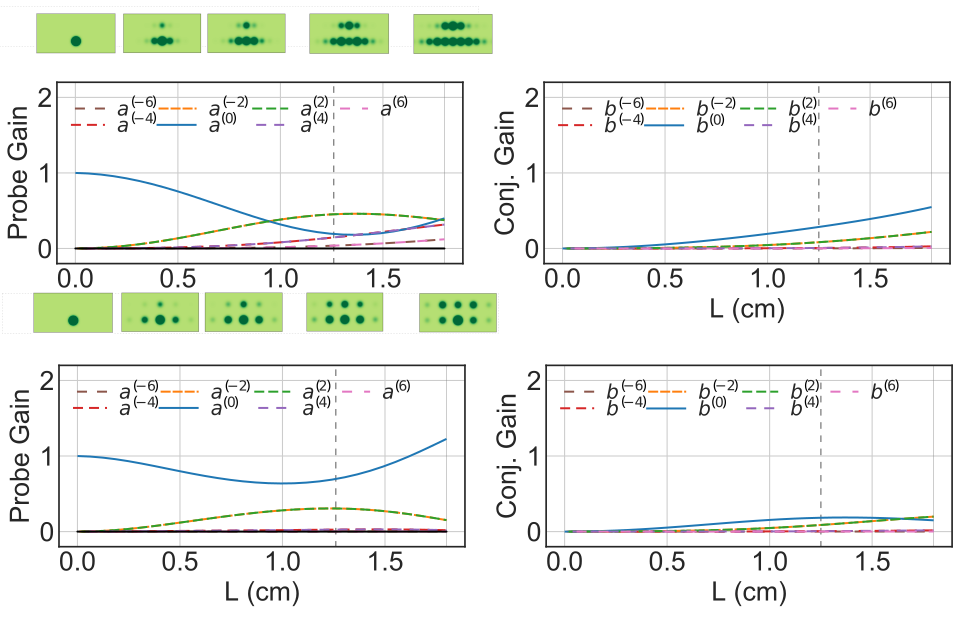}
\put(-2,56){(a)}
\put(-2,26){(b)}
\end{overpic}
\caption{Propagation of the modes for two different angles: (a) $\theta_{\mathrm{eff}}=3$~mrads and $\theta_{\mathrm{eff}}=6$~mrads. The parameters of the calculation are: $\Omega_0/2\pi=220$~MHz, $\Delta/2\pi=0.9$~GHz, $\Gamma/2\pi=5.7$~MHz,  $\gamma_d/2\pi=1$~MHz, $\omega_{\mathrm{HF}}/2\pi=3.035$~GHz and $g_a=g_b=0.28$~MHz.}
\label{fig:modes_propagation_d2_20MHz}
\end{figure}

Now, we look at the propagation of the multimode process when we blue detuned the seed beam far from the two-photon resonance, around $\delta/2\pi=20$~MHz in Fig.\ref{fig:modes_propagation_d2_20MHz}. In this case, for small angles in Fig.\ref{fig:modes_propagation_d2_20MHz}(a), the seed beam is deeply attenuated, resulting in the amplification of a greater number of modes. In contrast, for large angles in Fig.\ref{fig:modes_propagation_d2_20MHz}(b), there is robust propagation of the seed beam, with few modes generated.

\begin{figure}[b!]
\centering
\begin{overpic}[width=0.48\textwidth]{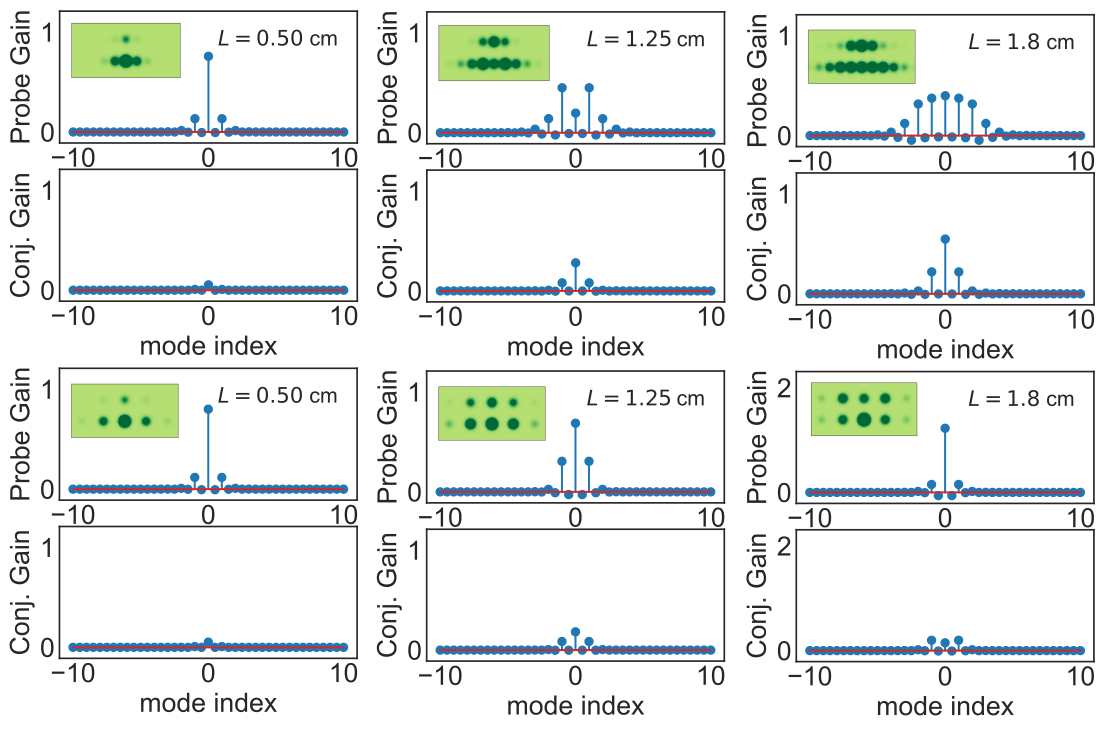}
\put(7,47){(a)}
\put(40,47){(b)}
\put(73,47){(c)}
\put(7,14){(d)}
\put(40,14){(e)}
\put(73,14){(f)}
\end{overpic}
\caption{Gain distribution for the probe and conjugate channel. (a) and (b) for $\theta_{\mathrm{eff}}=3$~mrads, (c) and (d) for $\theta_{\mathrm{eff}}=4.8$~mrads and (e) and (f) for $\theta_{\mathrm{eff}}=6$~mrads. The parameters of the calculation are: $\Omega_0/2\pi=220$~MHz, $\Delta/2\pi=0.9$~GHz, $\Gamma/2\pi=5.7$~MHz,  $\gamma_d/2\pi=1$~MHz, $\omega_{\mathrm{HF}}/2\pi=3.035$~GHz and $g_a=g_b=0.28$~MHz.}
\label{fig:modes_idx_propagation_d2_20MHz}
\end{figure}
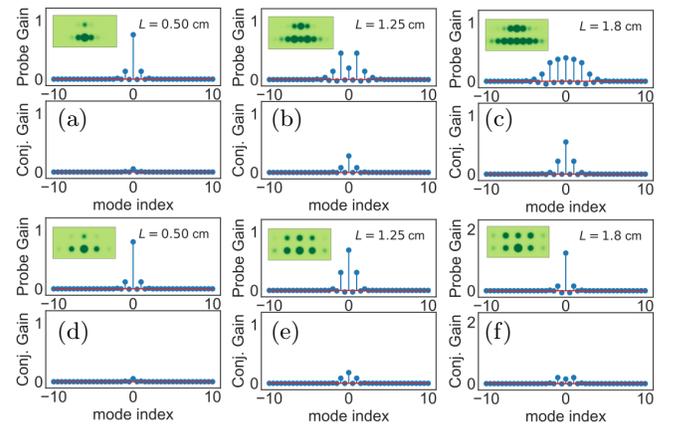

The gain distribution for 3 sections is plotted in Fig.~\ref{fig:modes_idx_propagation_d2_20MHz}. In this case, at $L=0.5$~cm, the gain distribution is also $3\times1$, with a weaker conjugate field than in the case of $\delta/2\pi=5.5$~MHz. Then, at $L=1.25$~cm, the distribution is $5\times3$ for small angles, while it is $3\times3$ for large angles.
However, in the case of $5\times3$, the amplification gain presents an inhomogeneous distribution in the probe channel. Interestingly, for a longer propagation of $L=1.8$~cm, the gain distribution in the probe channel is more homogeneous than in the case of $L=1.25$~cm. In the case of a large angle, the amplification in the conjugate channel is weaker compared to the other cases.
 
Overall, we can observe that, in either case, the seed beam is attenuated in order to excite the new modes. Nevertheless, the angle and the two-photon detuning determine the exponential amplification, as for $\delta/2\pi \sim 5.5$~MHz near the resonance  and small angle $\theta_{\mathrm{eff}}\sim 3$~mrads, while a weaker and more discrete multimode distribution occurs for off resonance and large angles, such as $\delta/2\pi \sim 20$~MHz and $\theta_{\mathrm{eff}}\sim 6$~mrads.

So far, we have analysed the gain distribution in terms of two-photon detuning, the angle between the pump fields, and the propagation through the atomic medium. This only gives us information about the classical variables of the multimode process. In the next section, we investigate the noise properties of the multiple fields generated in the system, which contain all the quantum properties of the multimode system.

\section{Noise properties of the multimode system\label{sec:Noise}}


As in the case of a single pump in Sec.~\ref{sec:4WM_sgl_pump}, in order to analyse the quantum properties of the multimode process, we employ the linearisation for the atomic and light operators to treat the quantum fluctuations in the frequency domain. This approach allows us to directly compare the solution with experimental observations.

\begin{figure}[h!]
\centering
\includegraphics[width=86mm]{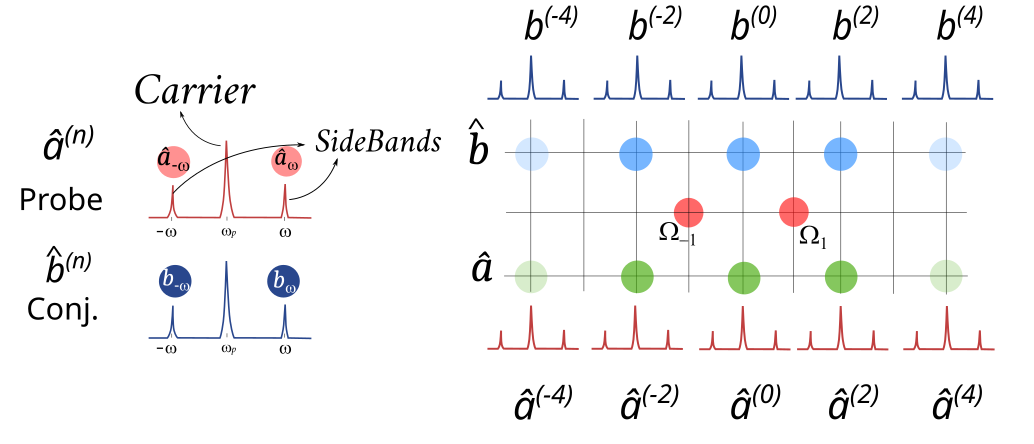}
\caption{Representation of sideband modes for the generated fields $\hat{a}^{(n)}$ and $\hat{b}^{(n)}$.}
\label{fig:M4WM_sidebands}
\end{figure}

The amplitude and intensity fluctuations of the optical fields can be spectrally decomposed by a carrier field with a central frequency, and the upper and lower sideband frequency modes, as represented in Fig.~\ref{fig:M4WM_sidebands}.  In the multimode scenario, each mode has its own pair of sideband modes. Each of these frequency modes can be described by canonical observables  $\hat{p}_{\pm\omega}^{(n)}=\hat{a}_{\pm\omega}^{(n)}+\hat{a}^{(n)\dagger}_{\pm\omega}$ and $\hat{q}_{\pm\omega}^{(n)}=-i[\hat{a}_{\pm\omega}^{(n)}-\hat{a}^{(n)\dagger}_{\pm\omega}]$, with $[\hat{a}_\omega^{(n)},\hat{a}^{(m)\dagger}_{\omega'}]=\delta_{nm}\delta(\omega-\omega')$. However, the transformation from the time domain to the frequency domain, using the Fourier transform, recombines the sideband modes such that the field quadratures from the Fourier transform $\delta\hat{p}(t)=\int d\omega \delta\hat{p}(\omega) e^{-i\omega\, t}$ can be written as $\hat{p}_\pm(\omega)=\hat{a}_\omega+\hat{a}^\dagger_{-\omega}$. In any case, determining $\hat{p}_\pm(\omega)$ or  $\hat{p}_{\pm\omega}$ allows us to reconstruct the state of each one of the fields.

\subsection{Noise power spectral density matrix}

\HM{}By applying the Fourier transform to the operators of each mode $\delta\hat{\mathbf{X}}(t)=\int d\omega \delta\hat{\mathbf{X}}(\omega) e^{-i\omega\, t}$, the vector of modes for the atomic operator is written as
\begin{align}
\delta\mathbb{\hat{X}}(\omega)=&\tilde{\mathbb{M}}(\omega)[ \mathbb{G}_x \delta\mathbb{\hat{A}}(\omega)+\mathbb{F}(\omega)  ],\label{eq:d_deltaXFloqFFT}
\end{align}
in which we defined $\tilde{\mathbb{M}}(\omega)=-[i\omega\mathbf{I}+\mathbb{M}]^{-1}$, with $\omega$ as the analysis frequency. The first term corresponds to the atomic response to the coupling with the fluctuations of the generated modes $\delta\mathbb{\hat{A}}(\omega)$, and the second term corresponds to the coupling with the stochastic operators from the vacuum. Thus, the optical field mode fluctuations follow the propagation equation
\begin{align}
\frac{\partial \delta \mathbb{\hat{A}}(\omega)}{\partial z}&=\mathbb{R}(\omega)\delta \mathbb{\hat{A}}(\omega)+\mathbb{R}_F(\omega)\mathbb{F}(\omega),\label{eq:d_deltaAFloqExp}
\end{align}
where we defined the coupling matrices as $
\mathbb{R}(\omega)=-\frac{N}{c}\mathbb{T}\, \tilde{\mathbb{M}}(\omega)\, \mathbb{G}_x +i(\omega/c) \mathbb{I}-i\mathbb{N}\, \Delta k_z z$ and $
\mathbb{R}_F(\omega)=-\frac{N}{c}\mathbb{T}\, \tilde{\mathbb{M}}(\omega)$.
This equation is equivalent to the expression for the single-pump case in eq.(\ref{eq:deltaA_omega}).
Hence, the solution for the propagation of the fields is given by
\begin{align}
\delta\mathbb{\hat{A}}(z,\omega)&=\mathbb{J}(z,\omega)
\delta\mathbb{\hat{A}}(0,\omega)+\mathbb{J}(z,\omega)
\mathbb{F}_{in}(z,\omega),\label{eq:SolFFTdA_n}
\end{align}
where we defined the propagator $\mathbb{J}(z,\omega)=e^{\mathbb{R}(\omega)\ z}$ and the stochastic operator vector is defined in the appendix~\ref{app:stc_noise}. Thus, this solution contains the amplitude fluctuation for each spatial mode $\delta \hat{a}^{(n)}$ and $\delta \hat{b}^{(n)}$, and each sideband mode at a particular analysis frequency $\omega$.

From this solution, we can obtain the power spectral density matrix from the amplitude field operators, which is defined by 
$\mathbb{S}(z,\omega)=(2\pi c/L)  \langle\delta\mathbb{\hat{A}}(z,\omega),\delta\mathbb{\hat{A}}(z,\omega')^T\rangle\, \delta(\omega+\omega')$.
Substituting eq.(\ref{eq:SolFFTdA_n}) into the equation above, we have
\begin{align}
\mathbb{S}(z,\omega)=&\left[ \mathbb{J}(z,\omega)\mathbb{S}(0,\omega)\mathbb{J}(z,\omega')^T\right.\nonumber\\
&+\left.\mathbb{J}(z,\omega)\mathbb{S}_F(z,\omega)\mathbb{J}(z,\omega')^T\right],
\end{align}
in which
 $\mathbb{S}(0,\omega)=(2\pi c/L)\langle\delta\mathbb{\hat{A}}(0,\omega),\delta\mathbb{\hat{A}}(0,\omega')^T\rangle\, \delta(\omega+\omega')
=2\pi \mathbb{S}_A(\omega)\, \delta(\omega+\omega')
$ corresponds to the noise power spectral density from the input field modes, and $\mathbb{S}_F(z,\omega)=(2 c/L)\Braket{\mathbb{\hat{F}}_{in}(z,\omega),\mathbb{\hat{F}}_{in}(z,\omega')^T}$ corresponds to the contribution from the stochastic terms. Hence, the elements $[\mathbb{S}(z,\omega)]_{n,-n}$ correspond to the noise spectral density matrix of each mode $(n)$. 

\subsection{Covariance Matrix}

In order to reconstruct the covariance matrix of set of modes that are amplified by the atoms, we define the generalised quadratures of the fields as $\hat{p}^{(n)}(\omega)=\hat{a}^{(n)}(\omega)e^{\phi^{(n)}}+\hat{a}^{(n)\dagger}(\omega) e^{-\phi^{(n)}}$ and $\hat{q}^{(n)}(\omega)=-i(\hat{a}^{(n)}(\omega) e^{\phi}-\hat{a}^{(n)\dagger} (\omega) e^{-\phi^{(n)}})$, where $\phi^{(n)}$ corresponds to the phase acquired by the carrier along the propagation through the medium. Now, in the matrix representation, we define the vector of quadratures fluctuations as  $\delta\mathbb{\hat{P}}(z,\omega)=\mathbb{U}(z)\delta\mathbb{\hat{A}}(z,\omega)$, and for each mode, we have
\begin{align}
\delta \mathbb{\hat{P}}^{(n)}(z,\omega)&=\begin{bmatrix}
  e^{i\phi_a^{(n)}} & e^{-i\phi_a^{(n)}} & 0 & 0\\
    -ie^{i\phi_a^{(n)}} & ie^{-i\phi_a^{(n)}} & 0&0\\
    0&0&e^{i\phi_b^{(n)}} & e^{-i\phi_b^{(n)}}\\
    0&0&-ie^{i\phi_b^{(n)}} & ie^{-i\phi_b^{(n)}}
\end{bmatrix}
\begin{bmatrix}
  \delta \hat{a}^{(n)}\\
  \delta \hat{a}^{ \dagger (n) }\\
    \delta \hat{b}^{(n)}\\
  \delta \hat{b}^{ \dagger(n)}
\end{bmatrix},
\end{align}
where we defined the vector $\delta \mathbb{\hat{P}}^{(n)}(z,\omega)=[\delta \hat{P}_a^{(n)}(z,\omega), \delta \hat{Q}_a^{(n)}(z,\omega),\delta \hat{P}_b^{(n)}(z,\omega),\delta \hat{Q}_b^{(n)}(z,\omega)]^T$~with amplitude and phase quadrature fluctuations $\delta \hat{P}_a^{(n)}(z,\omega)$ and $\delta \hat{Q}_a^{(n)}(z,\omega)$ for each mode of the probe and conjugate channel. Considering that each mode gains its own dephase $e^{i\phi^{(n)}}$ as it propagates through the medium, the  transformation considers each phase and makes the matrix $\mathbb{U}$ $z$-dependent, since the phases $e^{i\phi_a^{(n)}}$  and $e^{i\phi_b^{(n)}}$ are calculated at a particular position $z$, from eq.(\ref{eq:SolAHarm_z}). As a result, the covariance matrix $\mathbb{V}$ can be written as 
\begin{align}
\mathbb{V}(z,\omega)=& \mathbb{U}(z)\mathbb{J}(z,\omega)\mathbb{S}(0,\omega)\mathbb{J}(z,\omega')^T\mathbb{U}(z)^T\nonumber\\
&+\mathbb{U}(z)\mathbb{J}(z,\omega)\mathbb{S}_F(z,\omega)\mathbb{J}(z,\omega')^T\mathbb{U}(z)^T,
\end{align}

This is the most important result of the work, as it demonstrates the algebraic treatment from the initial state of the seed beam $\mathbb{S}(0,\omega)$ through the propagation in the medium $\mathbb{J}(z,\omega)$, until the covariance matrix at the output after the interaction with the atoms $\mathbb{V}(z,\omega)$. 

The detection of photocurrents to extract the noise at particular sideband modes involves the symmetric combination of fluctuations encoded in  $\omega$ and $\omega'=-\omega$. Therefore, by symmetrising the covariance matrix as $\mathbb{V}_S=1/2[\mathbb{V}(\omega)+\mathbb{V}(-\omega)]$, for a given number of total modes $d_H=2Q+1$, the form of the covariance matrix is

\begin{widetext}
\begin{eqnarray}
\mathbb{V}_S(z,\omega) &=& \scriptsize{\begin{pmatrix}
		\mathbf{V}^{(-Q)} & \mathbf{C}^{(-Q,-Q+1)} & \cdots & \cdots & \cdots & \cdots & \mathbf{C}^{(-Q,Q)} \\
		\cdots & \cdots & \cdots & \cdots & \cdots & \cdots & \cdots\\
		\cdots & \mathbf{C}^{(-1,-2)} & \mathbf{V}^{(-1)} & \mathbf{C}^{(-1,0)} & \cdots & \cdots & \cdots \\
		\cdots & \cdots & \mathbf{C}^{(0,-1)} & \mathbf{V}^{(0)} & \mathbf{C}^{(0,1)} & \cdots & \cdots \\
        \cdots & \cdots & \cdots & \mathbf{C}^{(1,0)} & \mathbf{V}^{(1)} & \mathbf{C}^{(1,2)} & \cdots \\
        \cdots & \cdots & \cdots & \cdots & \cdots & \cdots & \cdots \\
        \mathbf{C}^{(Q,-Q)} & \cdots & \cdots & \cdots & \cdots & \mathbf{C}^{(Q,Q-1)} & \mathbf{V}^{(Q)} \\
	\end{pmatrix}}\ \label{eq:Vs}
\end{eqnarray}
\end{widetext}
where the diagonal matrices $\mathbf{V}^{(n)}$ correspond to the covariance matrix for the $(n)$-mode for the probe and conjugate channels, such that
\begin{align}
\scriptsize{\mathbf{V}^{(n)}(z,\omega)} &=\scriptsize{\begin{bmatrix}
		\Delta^2 \p_a^{(n)} & \braket{\p_a^{(n)},\q_a^{(n)} } & \braket{\p_a^{(n)},\p_b^{(n)} } & \braket{\p_a^{(n)},\q_b^{(n)} }\\
		  \braket{\q_a^{(n)},\p_a^{(n)} } & \Delta^2 \q_a^{(n)} & \braket{\q_a^{(n)},\p_b^{(n)} } & \braket{\q_a^{(n)},\q_b^{(n)}}\\
		 \braket{\p_b^{(n)},\p_a^{(n)} } &  \braket{\p_b^{(n)},\q_a^{(n)} } & \Delta^2 \p_b^{(n)} & \braket{\p_b^{(n)},\q_b^{(n)} }\\
         \braket{\q_b^{(n)},\p_a^{(n)} } &  \braket{\q_b^{(n)},\q_a^{(n)} } & \braket{\q_b^{(n)},\q_b^{(n)} }  & \Delta^2 \q_b^{(n)}
	\end{bmatrix}},\label{eq:V_n}
\end{align}
while the off-diagonal matrices $\mathbf{C}^{(n,m)}$ contain all the correlations among the mode $(n,m)$ with $n,m=-Q,-Q+1,\cdots Q-1,Q$.

A final important aspect to consider is the correction by detection efficiency. To include this, we introduce a beam splitter transformation to obtain the effective covariance matrix, as explained in appendix~\ref{app:losses}. Throughout the paper, a detection efficiency of $\eta=0.95$ is considered, as reported in ref.~\cite{Jing2020}. 
\subsection{Multimode Quantum correlations spectrum}

A direct way to observe the quantum correlations among the generated modes is to look at the noise difference (ND) between two given modes. In particular, for any pair of modes $\p_\alpha^{(n)}$ and $\p_\beta^{(m)}$, with $\alpha=a,b$  and $n,m=0,\pm 1,\pm2,\dots,\pm Q$, the noise difference is given by
\begin{align}
\Delta^2 \hat{P}_-^{(n,m)}(z,\omega)=&\Delta^2 \hat{P}_\alpha^{(n)}(z,\omega) + \Delta^2 \hat{P}_\beta^{(m)}(z,\omega)\nonumber\\
&-\braket{\p_\alpha^{(n)},\p_\beta^{(m)}}-\braket{\p_\beta^{(m)},\p_\alpha^{(n)}},
\end{align}
such that if $\Delta^2 \hat{P}_-^{(n,m)}(z,\omega)<2$, the pair of spatial modes present quantum correlations. Nevertheless, for a multimode state, it is expected to have thermal states for any bipartition such that $\Delta^2 \hat{P}_-^{(n,m)}(z,\omega)\geq 2$.

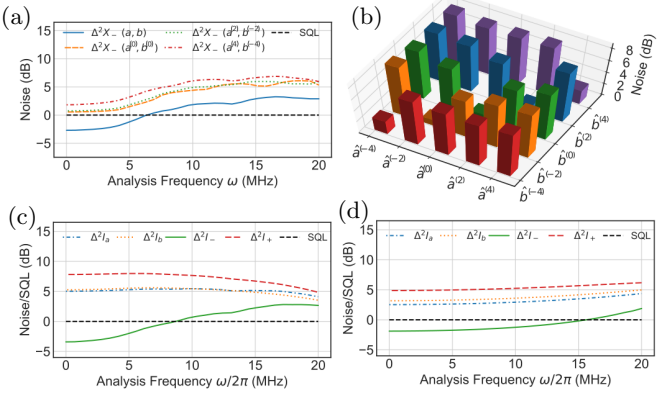
\begin{figure}[t!]
\centering
\begin{overpic}[width=0.48\textwidth]{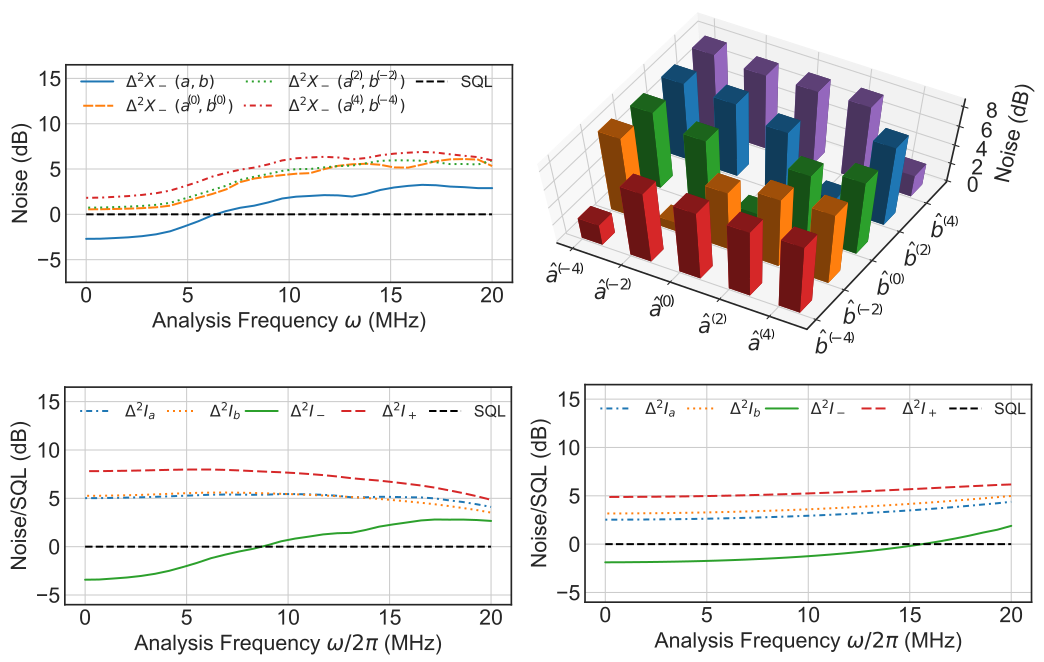}
\put(-2,57){(a)}
\put(52,57){(b)}
\put(-1,26){(c)}
\put(50,27){(d)}
\end{overpic}
\caption{(a) Noise difference spectrum for bipartitions $(\hat{a}^{(0)},\hat{b}^{(0)})$(orange dashed line), $(\hat{a}^{(2)},\hat{b}^{(-2)})$(green dotted line), $(\hat{a}^{(4)},\hat{b}^{(-4)})$(red dashed line) and the two mode quantum limit (SQL). The total noise difference is plotted in blue solid line. (b) Noise difference for each pair of modes. The intensity noise spectrum for each channel probe $(\Delta^2 \hat{I}_a)$ and conjugate $(\Delta^2 \hat{I}_b)$, and the sum and difference  $(\Delta^2 \hat{I}_+)$ and  $(\Delta^2 \hat{I}_-)$, are plotted in  (c) and (d) for near the resonance $\delta/2\pi=5.5$~MHz and far of the resonance $\delta/2\pi=20$~MHz, respectively.}
\label{fig:multi_partite_corr}
\end{figure}

Figure~\ref{fig:multi_partite_corr}(a) shows the ND for symmetric pairs of modes as a function of the analysis frequency. Notice that for all of them, the ND shows an excess of noise such that $\Delta^2 \hat{P}_-^{(n,m)}(z,\omega)>2$. However, the total noise difference, which is determined by $\Delta^2\hat{P}_-=\Delta^2(\sum_{n=-2}^{2}[\p_a^{(2n)}-\p_b^{(-2n)}])$, presents squeezing for $\omega/2\pi<7$~MHz such that $\Delta^2\hat{P}_-<d_H$ with $d_H=10$ as the total number of modes.

In addition to that, we test all possible bipartitions for the case of $5\times5$ modes. Figure~\ref{fig:multi_partite_corr}(b) shows the ND of all the bipartitions for $\omega/2\pi=2$~MHz, within the bandwidth of squeezing, where the total difference presents squeezing in figure (a). The histogram shows that indeed, all the bipartitions present excess noise. As in ref.~\cite{Jing2018}, this is a first indicator that the M4WM produces a quantum correlated multimode state.

Furthermore, most of the experimental implementations measured the noise intensity difference (NID) in squeezing to demonstrate the quantum properties of multimode generation. The NID can be calculated from the density matrix and the gain of the modes  in eq.(\ref{eq:gain_modes_f}). To do so, the intensity fluctuation of any of the modes $\delta \hat{I}^{(n)}(z,\omega)$ is proportional to its amplitude quadrature fluctuations $\delta \p^{(n)}(z,\omega)$, such that for both channels $\hat{a}$ and $\hat{b}$ they are defined as 
\begin{align}
\delta \hat{I}_a^{(n)}(z,\omega)=|\alpha^{(n)}|\delta \p_a^{(n)}(z,\omega)\\
\delta \hat{I}_b^{(n)}(z,\omega)=|\beta^{(n)}|\delta \p_b^{(n)}(z,\omega)
\end{align}
where $|\alpha^{(n)}|$ and $|\beta^{(n)}|$ are obtained from the gain solution in eq.(\ref{eq:gain_modes_f}). For any pair of modes $(n,m)$ in any channel $a,b$, the fluctuations of the sum and the difference are defined as $\delta \hat{I}_{\pm}^{(n,m)}(z,\omega)=|\alpha^{(n)}|\delta \hat{X}_a^{(n)}(z,\omega)~\pm~|\beta^{(m)}|\delta \hat{X}_b^{(m)}(z,\omega)$. As in the case of amplitudes, when the two-mode intensity difference $\Delta^2\hat{I}_-=\Delta^2(\sum_{n=-2}^{2}[\delta \hat{I}_a^{(2n)}-\delta \hat{I}_b^{(-2n)}])$ presents squeezing $\Delta^2\hat{I}_-<2$, the two channels share intensity quantum correlations.

Figure~\ref{fig:multi_partite_corr}(c) shows the noise spectra for the bipartition $(\hat{a},\hat{b})$, comparing the noise of each channel $\Delta^2 I_a$ and $\Delta^2 I_b$, the noise of the sum $\Delta^2 I_+$ and the NID $\Delta^2 I_-$. For this case, we consider an angle of $\theta_{\mathrm{eff}}=3$~mrad and a two-photon detuning of $\delta/2\pi=5.5$~MHz.  The noise spectra for each field and the sum exhibit an excess of noise of  5 and 10 dB, respectively, whereas the NID presents a  maximum squeezing level of -4~dB. This is expected from a two-mode quantum correlated system: thermal states for each subsystem while noise compression in the NID exhibits quantum correlations. In addition to that, the noise spectra show a squeezing bandwidth of around $\omega/2\pi\sim 9$~MHz. For $\omega/2\pi> 9$~MHz, the total NID presents an excess of noise, so the modes in the probe and conjugate channels are not quantum correlated in amplitude. The  bandwidth of squeezing is mainly determined by the Rabi frequency of the pump field and the two photon detuning. However, in this multimode process, as we will show next, it also depends on the angle between the two pump fields. Nevertheless,  this result for the IDS of the pairs of modes and the IDS squeezing spectrum of the bipartition is consistent with the  observations in ref.\cite{Jing2018}.

In the case above, we investigate the squeezing bandwidth when the seed beam is near the two-photon resonance. Now, we can observe how the squeezing bandwidth changes with the seed beam blue detuned in the two-photon resonance. Figure~\ref{fig:multi_partite_corr}(d) plots the noise spectrum for the case of $\delta/2\pi=20$~MHz.
In this case, the total NID squeezing level is reduced,  but the bandwidth is increased. This is consistent with the case of a single pump, where the NID squeezing bandwidth increases as the two-photon detuning increases, at the cost of reducing the squeezing level.

\begin{figure}[b!]
\centering
\begin{overpic}[width=86mm]{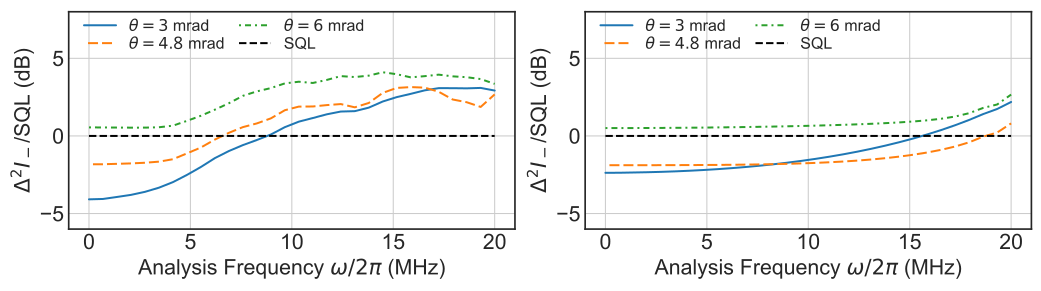}
\put(-1,26){(a)}
\put(50,26){(b)}
\end{overpic}
\caption{Noise Intensity difference spectrum for $\theta_{\mathrm{eff}}=3$~mrads, $\theta_{\mathrm{eff}}=4.8$~mrads and $\theta_{\mathrm{eff}}=6$~mrads when the seed beam is tuned at (a) $\delta/2\pi=5.5$~MHz and (b) $\delta/2\pi=20$~MHz. The parameters of the calculation are: $\Omega_0/2\pi=220$~MHz, $\Delta/2\pi=0.9$~GHz, $\Gamma/2\pi=5.7$~MHz,  $\gamma_d/2\pi=1$~MHz, $\omega_{\mathrm{HF}}/2\pi=3.035$~GHz and $g_a=g_b=0.28$~MHz. Detection efficiency $\eta=0.95$.}
\label{fig:multi_partite_corr_angle}
\end{figure}

Since the total NID witnesses the quantum correlation of the multimode state, we analyse its behaviour for different angles. Figure~\ref{fig:multi_partite_corr_angle}(a) and (b) show the NID spectrum for 2 different photon detunings, respectively. In each case, we plot the NID for 3 different angles of the pump fields. We can observe that for a low angle $\theta_{\mathrm{eff}}=3$~mrad and near the two-photon resonance $\delta/2\pi=5.5$~MHz, the NID in Fig.~\ref{fig:multi_partite_corr_angle}(a) reaches a squeezing level of $-4$dB for low analysis frequency and a bandwidth of $\omega/2\pi\sim9$~MHz. As the angle is increased to $\theta_{\mathrm{eff}}=4.8$~mrad, the squeezing level is reduced to $-2.4$dB and the bandwidth decreases to $\omega/2\pi\sim7$~MHz. Now, if the angle is further increased to $\theta_{\mathrm{eff}}=6$~mrad, the quantum correlations in amplitude are lost with an excess of noise.

A relatively similar behaviour can be observed in Fig.~\ref{fig:multi_partite_corr_angle}(b). The overall bandwidth is increased; the NDI for the two angles $\theta_{\mathrm{eff}}=3$~mrad and $\theta_{\mathrm{eff}}=4.8$~mrad preserves the squeezing; however,  for $\theta_{\mathrm{eff}}=6$~mrad, the NID presents an excess of noise. It is worth mentioning that, although the squeezing is lost as the angle is increased, properties such as entanglement might still be preserved in the system. In the following section, we investigate the entanglement of the multimode system under different situations.

\subsection{Two-photon detuning spectroscopy of the  Multimode Quantum correlations}

In ref.\cite{Jing2019}, the authors show that the NID squeezing is observed for blue two photon detuning, whereas for red two-photon detuning, the NID presents an excess of noise. In this section, we aim to briefly show that the model reproduces this kind of behaviour.

\begin{center}
\begin{figure}[h!]
\begin{overpic}[width=\textwidth]{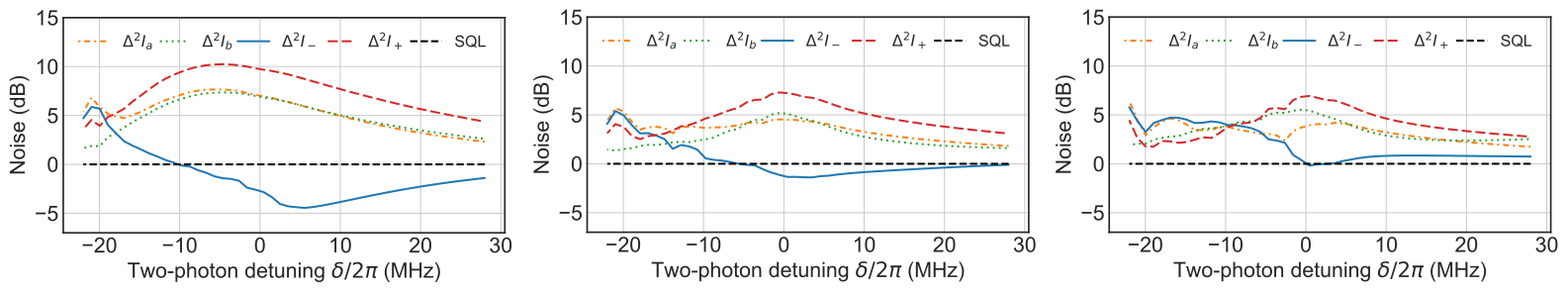}
\put(-1,16){(a)}
\put(33,16){(b)}
\put(66,16){(c)}
\end{overpic}
\caption{Intensity noise two-photon spectroscopy for each channel probe $(\Delta^2 \hat{I}_a)$ and conjugate $(\Delta^2 \hat{I}_b)$, and the sum and difference  $(\Delta^2 \hat{I}_+)$ and  $(\Delta^2 \hat{I}_-)$, for  (a) $\theta_{\mathrm{eff}}=3$~mrads, (b) $\theta_{\mathrm{eff}}=4.8$~mrads and (c) $\theta_{\mathrm{eff}}=6$~mrads. The parameters of the calculation are: $\omega/2\pi=2$~MHz, $\Omega_0/2\pi=220$~MHz, $\Delta/2\pi=0.9$~GHz, $\Gamma/2\pi=5.7$~MHz,  $\gamma_d/2\pi=1$~MHz, $\omega_{\mathrm{HF}}/2\pi=3.035$~GHz and $g_a=g_b=0.28$~MHz. Detection efficiency $\eta=0.95$.}
\label{fig:Noise_vs_d2_var_angle}
\end{figure}
\end{center}

Therefore, we choose $\omega/2\pi=2$~MHz to investigate its behaviour as a function of the two-photon detuning $\delta$, as is done in ref.\cite{Jing2019}. We choose low frequencies from the noise spectrum profile since the  maximum squeezing of the total NID occurs at low analysis frequencies. Figure~\ref{fig:Noise_vs_d2_var_angle} shows the intensity noise as a function of the two-photon detuning for the 3 different angles of the pump fields. 
\onecolumngrid
\twocolumngrid
In the case of the angle $\theta_{\mathrm{eff}}=3$~mrad, the maximum  squeezing level is reached for $\delta/2\pi\sim5.5$~MHz, and it is preserved for $\delta/2\pi>-8$~MHz.
In contrast, the NID presents an excess of noise  for $\delta/2\pi<8$~MHz, which indicates the loss of quantum correlations. Thus, the profile within that range of frequencies is similar to the observation in ref.\cite{Jing2019}.

Conversely, for an angle of $\theta_{\mathrm{eff}}=4.8$~mrad, the maximum squeezing level is reduced to $-1$dB and is completely lost for $\theta_{\mathrm{eff}}=6$~mrad. This feature is also observed experimentally; relatively low angles lead to the maximum squeezing level, whereas the maximum squeezing level is reduced as the angle increases.

It is worth noting that the analysis presented above for the noise of the quadratures and the intensity fluctuations aims to show the bandwidth of the quantum properties in the  multimode process and how it behaves around the two photon detuning, which can be directly mapped to what is measured in typical experiments of this kind. Also, in the three cases, the multimode system exhibits  the two-mode properties, with excess noise for each channel as well as excess noise of the sum, and noise compression for the NID under certain parameters of interaction, otherwise, it is simply a thermal state.  In the next section, we analyse the entanglement properties of the multimode process and determine under which conditions multimode entanglement is present in the system. 

\section{Multipartite entanglement \label{sec:Entanglement}}

The multimode correlations cannot capture the actual entanglement properties of multimode 4WM. In this section, we employ the criterion of positive under partial transposition (PPT)~\cite{Simon00} to determine the entanglement among the main bipartitions. To do so, we calculate the symplectic eigenvalues for a given bipartition with the partial transposition, and if they are lower than 1, the fields are entangled, otherwise, it is a separable bipartition.

\begin{center}
\begin{figure}[h!]
\begin{overpic}[width=\textwidth]{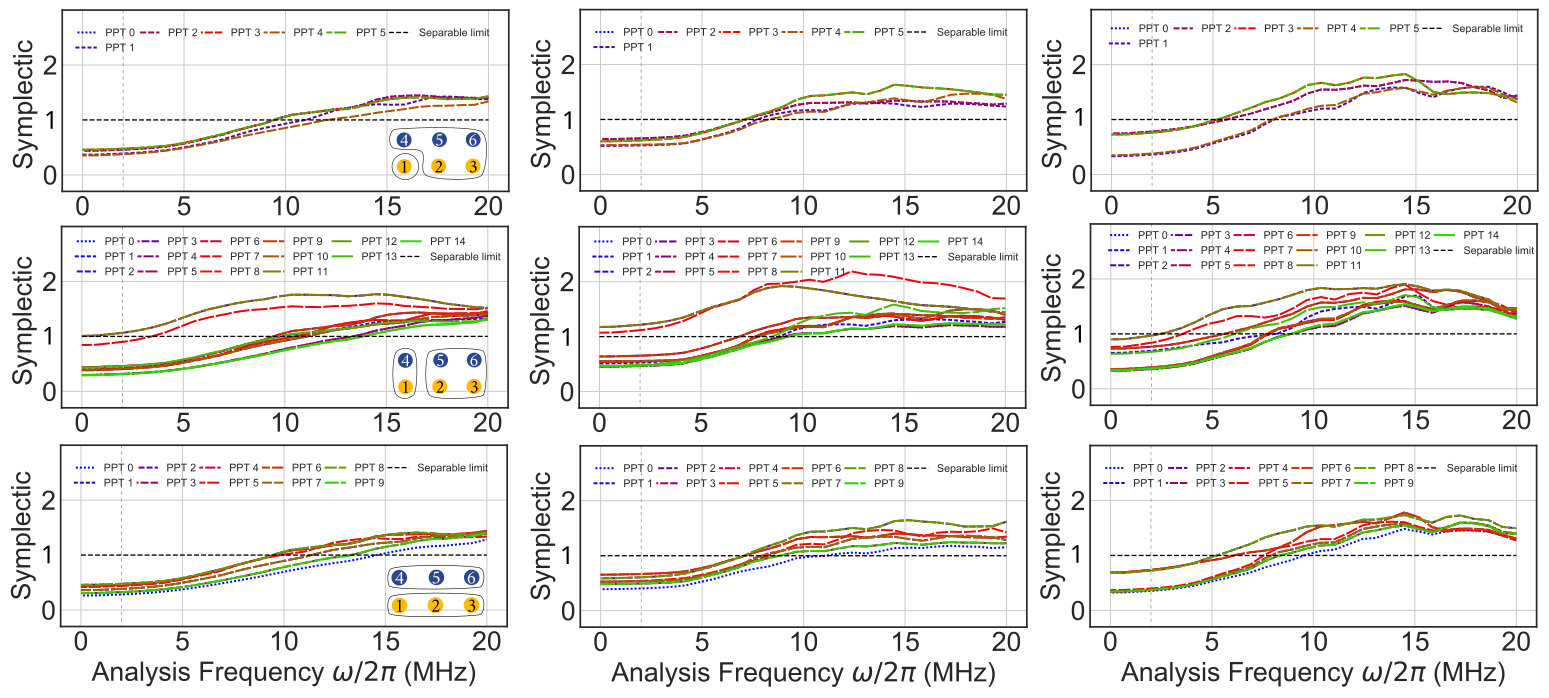}
\put(17,-1){(a)}
\put(50,-1){(b)}
\put(84,-1){(c)}
\end{overpic}
\caption{Simplectic eigenvalues spectrum for all 31 bipartitions of the hexaparite state. Figures at the top, middle and bottom location correspond to 6 bipartitions of $1\times 5$, 15 bipartitions of $2\times 4$ and 10 bipartitions of $3\times 3$, respectively. Figures (a), (b) and (c) correspond to (a) $\theta_{\mathrm{eff}}=3$~mrads, (b) $\theta_{\mathrm{eff}}=4.8$~mrads and (c) $\theta_{\mathrm{eff}}=6$~mrads, respectively. The parameters of the calculation: $\Omega_0/2\pi=220$~MHz, $\Delta/2\pi=0.9$~GHz, $\Gamma/2\pi=5.7$~MHz,  $\gamma_d/2\pi=1$~MHz, $\omega_{\mathrm{HF}}/2\pi=3.035$~GHz and $g_a=g_b=0.28$~MHz. Detection efficiency $\eta=0.95$.}
\label{fig:PPT_vs_omega_d2_5.5MHz}
\end{figure}
\end{center}

Let us start analysing the bandwidth of the  PPT criterion for different angles and the two-photon detuning of the seed beam. 
Since the number of possible bipartitions grows as the number of modes increases, we start analysing the case of 6 modes with a total number of bipartitions of 31. Figure~\ref{fig:PPT_vs_omega_d2_5.5MHz} shows the symplectic eigenvalues of PPT as a function of the analysis frequency for the three different angles used in the sections above and near the two-photon resonance $\delta/2\pi=5.5$~MHz. 
The three Figs.~\ref{fig:PPT_vs_omega_d2_5.5MHz} (a), (b), and (c), organised in columns, correspond to the PPT for $\theta_{\mathrm{eff}}=3$~mrads,$\theta_{\mathrm{eff}}=4.8$~mrads, and $\theta_{\mathrm{eff}}=6$~mrads, respectively. 
In each column, we present the PPT for three groups of bipartitions of the hexapartite scenarios: $1\times5$ in the top figure, $2\times4$ in the middle figure, and $3\times3$ in the bottom figure.


We begin with the first angle $\theta_{\mathrm{eff}}=3$~mrads in Fig.~\ref{fig:PPT_vs_omega_d2_5.5MHz} (a). The top figure shows that the 6 bipartitions $1\times5$ present symplectic values lower than 1 for $\omega/2\pi<10$~MHz. Therefore, any combination of modes $(1|2345)$ is entangled within a bandwidth of approximately 10~MHz, they are separable modes. In the middle figure, we can observe a relatively different scenario for bipartition  $2\times4$. The bipartitions $(16|2345)$ and $(34|1256)$ are not entangled within that range of frequencies, and the bipartition $(25|1436)$ presents a low bandwidth of entanglement up to $\omega/2\pi\sim3$~MHz. All other combinations of $2\times4$ bipartitions present a broader bandwidth of entanglement within a range of $\omega/2\pi\sim(9,14)$~MHz.  In the bottom figure, we can observe that the PPT for 10 bipartitions of $3\times3$ presents a behaviour similar to that of the $1\times5$ bipartitions in the top figure, with a similar bandwidth of entanglement. Notice that in this case, the bipartition $(123|456)$ (blue dotted line) achieves the lowest symplectic eigenvalue and thus, determines the lower limit of entanglement for that kind of bipartition. In other words, the modes on the probe channel against the modes of the conjugate channel are maximally entangled over a larger bandwidth, as in the single pump case.

\onecolumngrid
\twocolumngrid

Now we can analyse the results for $\theta_{\mathrm{eff}}=4.8$~mrads in Fig.~\ref{fig:PPT_vs_omega_d2_5.5MHz} (b). Overall, notice that the bandwidth of entanglement for those bipartitions $1\times5$, $2\times4$, and $3\times3$, with symplectic eigenvalues below 1, is reduced to $\omega/2\pi\sim7$~MHz. However, in this case, the bipartitions $(16|2345)$, $(34|1256)$, and $(25|1436)$ are not entangled in the 20~MHz frequency range. Thus,  as we increase the angle,  more bipartitions are separable in the hexapartite system.

In the last case, we now look at the results for $\theta_{\mathrm{eff}}=6$~mrads in  Fig.~\ref{fig:PPT_vs_omega_d2_5.5MHz} (c). Interestingly, the bandwidth of entanglement is reduced, but for low frequencies up to $\omega/2\pi \sim2$~MHz, all  bipartitions $1\times5$, $2\times4$, and $3\times3$ (from top to bottom) present entanglement. Therefore, this demonstrates that for large angles, the bandwidth is reduced, but the system shows hexapartite entanglement, i.e., all the bipartitions are entangled~\cite{Jing2020}. As the angle decreases, as in the case of $\theta_{\mathrm{eff}}=4.8$~mrads and $\theta_{\mathrm{eff}}=3$~mrads , the bandwidth of entanglement enhances for some bipartitions,  but the system does not present hexapartite entanglement. This is because, as the angle is reduced, more modes are generated, and the correlations are spread over a larger number of modes, so  we have to analyse a greater set of modes. Nevertheless, this result is consistent with what is observed in ref.~\cite{Jing2020}, where they show hexapartite entanglement exactly under those conditions: large angles and low analysis frequency. 

\subsection{Two-photon spectroscopy of the multimode entanglement}

In this last section, we analyse how the multimode entanglement of the hexapartite system, which was observed for sidebands with low frequencies, behaves for different values of two-photon detuning $\delta$. The tunability of these kinds of properties is of great relevance, and the atom-light interaction allows us to optically control such properties.

\onecolumngrid
\begin{center}
\begin{figure}[t!]
\begin{overpic}[width=\textwidth]{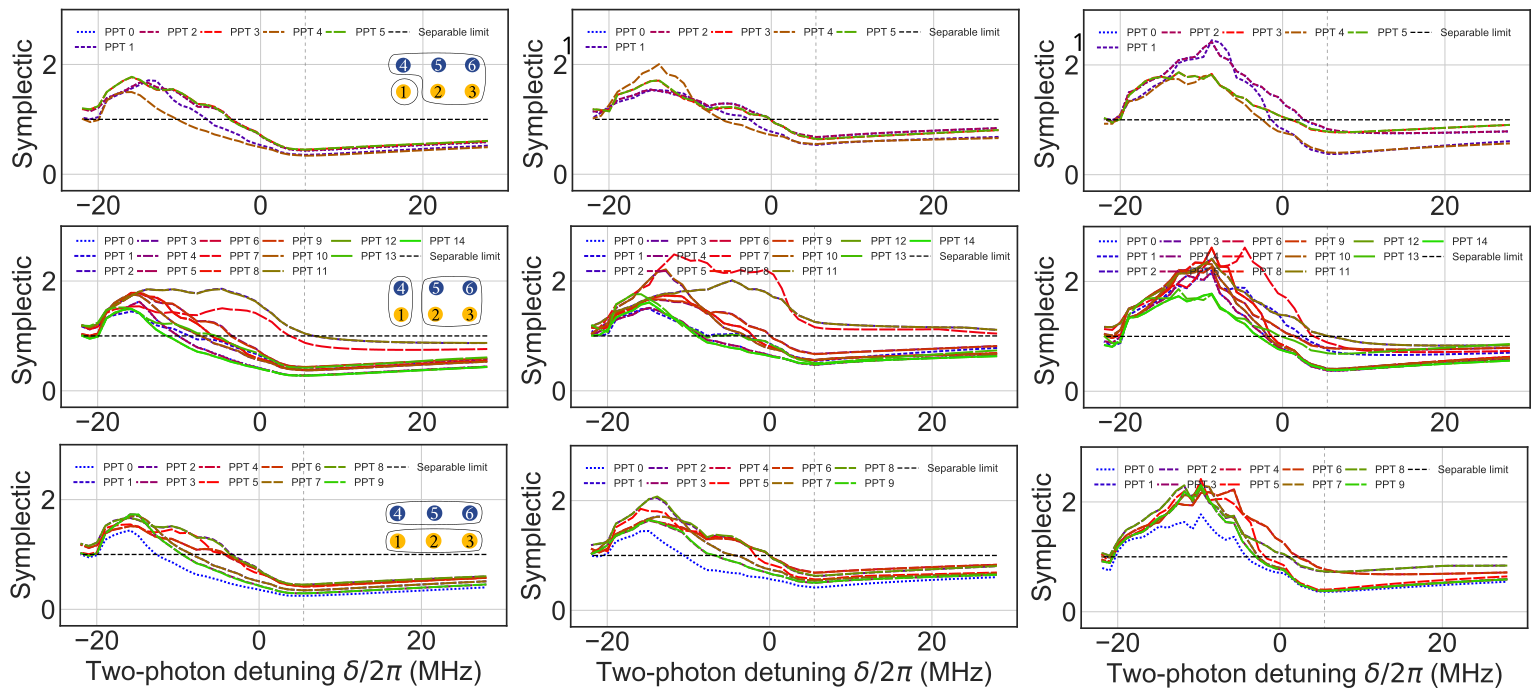}
\put(17,-1){(a)}
\put(50,-1){(b)}
\put(84,-1){(c)}
\end{overpic}
\caption{Symplectic eigenvalues two-photon spectroscopy for all 31 bipartitions of the hexapartite state. Figures at the top, middle and bottom correspond to 6 bipartitions of $1\times 5$, 15 bipartitions of $2\times 4$ and 10 bipartitions of $3\times 3$, respectively. Figures (a), (b) and (c) correspond to (a) $\theta_{\mathrm{eff}}=3$~mrads, (b) $\theta_{\mathrm{eff}}=4.8$~mrads and (c) $\theta_{\mathrm{eff}}=6$~mrads, respectively. The parameters of the calculation: $\omega/2\pi=2$~MHz, $\Omega_0/2\pi=220$~MHz, $\Delta/2\pi=0.9$~GHz, $\Gamma/2\pi=5.7$~MHz,  $\gamma_d/2\pi=1$~MHz, $\omega_{\mathrm{HF}}/2\pi=3.035$~GHz and $g_a=g_b=0.28$~MHz. Detection efficiency $\eta=0.95$.}
\label{fig:PPT_vs_d2_omega_2MHz}
\end{figure}
\end{center}
\twocolumngrid
Figure~\ref{fig:PPT_vs_d2_omega_2MHz} presents the symplectic eigenvalues for $\omega/2\pi=2$MHz as a function of the two-photon detuning for the three different angles used in the sections above. Figures~\ref{fig:PPT_vs_d2_omega_2MHz} (a), (b), and (c) are organised in columns corresponding to the PPT for $\theta_{\mathrm{eff}}=3$~mrads,$\theta_{\mathrm{eff}}=4.8$~mrads, and $\theta_{\mathrm{eff}}=6$~mrads, respectively. In each column, we present the PPT for three groups of bipartitions, as in the case of Fig.~\ref{fig:PPT_vs_omega_d2_5.5MHz}.

In the first angle $\theta_{\mathrm{eff}}=3$~mrads in Fig.~\ref{fig:PPT_vs_d2_omega_2MHz} (a),  all the bipartitions $1\times5$, $2\times4$ and $3\times3$, present symplectic eigenvalues lower than 1 for $\delta/2\pi>8$~MHz. In contrast, the bipartitions $(16|2345)$ and $(34|1256)$ (middle figure) are not entangled for $\delta/2\pi<6$~MHz, and  the bipartition $(25|1436)$ becomes separable for $\delta/2\pi<3$~MHz. Therefore, the system presents hexapartite entanglement when the seed beam is blue detuned from $\delta/2\pi>8$~MHz. It is interesting to note that, even though for $\delta/2\pi>5.5$~MHz, where the gain is maximal across all modes, the system is not fully entangled because they share correlations with other modes, which is typical for a multimode state. However, by detuning the seed beam far enough, the 6 modes become entangled, sharing strong correlations, independent of the presence of other modes.

A different situation occurs in Fig.~\ref{fig:PPT_vs_d2_omega_2MHz} (b) for $\theta_{\mathrm{eff}}=4.8$~mrads. Most of the bipartitions are entangled for $\delta/2\pi >0$~MHz  and are separable for $\delta/2\pi<0$~MHz. However, in contrast to the case above, the bipartitions $(16|2345)$, $(34|1256)$, and  $(25|1436)$ in the middle figure are separable for all values of $\delta$. Thus, even though there are more than 6 modes sharing correlations for this angle, as in the case above of $\theta_{\mathrm{eff}}=3$~mrads, there is no two-photon detuning to make the hexapartite system entangled.

Now, we analyse the case of $\theta_{\mathrm{eff}}=6$~mrads in Fig.~\ref{fig:PPT_vs_d2_omega_2MHz} (c). In this situation, we can observe that all the bipartitions are entangled for $\delta/2\pi>5.5$MHz, and therefore, we can obtain a hexapartite entangled state. On the other hand, for the red detuned seed beam for $\delta/2\pi<5$~MHz, the system begins to have separable bipartitions. Comparing with the other two cases, we can observe that as the angle increases, the system goes from a situation of hexapartite entanglement only at $\delta/2\pi>8$~MHz for small angles, through an intermediate angle where the system has no hexapartite entanglement, until a large angle where all the information is concentrated in the 6 modes, and therefore presents hexapartite entanglement for $\delta/2\pi>5.5$~MHz.

We can also observe that for bipartitions such as $1\times5$ and $3\times3$, the bandwidth of two-photon detuning is shifted for entangled partitions as the angle increases; i.e, the symplectic eigenvalues are below 1  for $\delta/2\pi>-5$~MHz when $\theta_{\mathrm{eff}}=3$~mrads, then it is shifted to  $\delta/2\pi>0$~MHz when $ \theta_{\mathrm{eff}}=4.8$~mrads, and subsequently to $\delta/2\pi>5.5$~MHz for $\theta_{\mathrm{eff}}=6$~mrads.

Therefore, the microscopic model for the multimode process offers a robust method to describe the gain distribution of the generated modes, the output multimode quantum state through the covariance matrix, and how these properties change with the parameters of interaction.

\section{Conclusion\label{sec:Conclusions}}
In conclusion, we have presented a comprehensive theoretical description of multimode generation based on a microscopic model of a double lambda system for the four-wave mixing process with multiple pumps. The work shows that the algebraic solution for the multiple pump case takes exactly the same form as that for the single pump case. This algebraic simplification of the multimode case allows us to describe the multimode generation in terms of the main parameters of interaction, such as light detuning with respect to the atomic levels, Rabi frequency, and the angle between the pump beams. The model we are presenting also shows a straightforward method to determine the covariance matrix in the frequency domain, which helps to predict the reconstruction of the multimode gaussian state after the interaction with the atoms. The results present satisfactory solutions for gain distribution, quantum correlations, and multimode entanglement for different values of two-photon detuning, analysis frequency, and pump angles. 

This kind of model is compatible not only with the 4WM, but it can also be exploited to investigate spectral control with 6WM in multimode states in alkali atoms, as in ref.\cite{Hans2025_1}. The model also offers the possibility of studying pump shaping and multiple pump fields without changing the form of the solutions. The model can also be extended to multimode transverse modes to explore reconfigurable quantum networks. Therefore, the robustness of the model can help in investigating the best tunability of the quantum properties of the multimode system, while the phenomenological approach cannot provide a physical insight of such a complex non-linear process.

\appendix

\section{Doppler integration \label{app:Doppler} }
In order to account for the Doppler broadening of the gain profile, we consider the Doppler shift $\omega_i-\mathbf{k}\cdot\mathbf{v}$, for each light field $\omega_i=(\omega_0,\omega_0^a,\omega_0^b)$, due to the velocity of the atoms. Here, we consider the $z$ component of $|\mathbf{k}_0|= k\cos{\theta/2}$ for the pump and $|\mathbf{k}^{a(n)}|=|\mathbf{k}^{b(n)}|=k\cos{n\theta/2}$ for the mode in the probe and conjugate channel. Hence, the vector of atomic operators $\mathbb{X}_{v_z}$ is now described for atoms with velocity $v_z$. 
Thus, the dynamical equation for a group of atoms with a velocity $\mathbf{v}$ is as follows
 \begin{align}
\frac{d\mathbb{\hat{X}}_{v_z}(z,t)}{dt}=&\mathbf{M}(v_z)
\mathbb{\hat{X}}_{v_z}(z,t)+ \mathbb{G}_{x} \mathbb{\hat{A}}(z,t)+\mathcal{\hat{F}}(z,t)_{v_z},\label{eq:Dyn5Level_v}
\end{align}

From the atomic operators contained in $\mathbf{X}_{v_z}$ for a given velocity $v_z$, we can now determine the propagation of the modes in the probe  and conjugate channels $\hat{a}^{(n)}(z,t)$ and $\hat{b}^{(n)}(z,t)$ 
\begin{align}
(\partial_t + c\partial_z)\hat{\mathbb{A}}(z,t)=\sum_v N\mathbb{T}\mathbb{\hat{X}}_{v_z}(z,t) \Delta v_z f(v_z) \label{eq:Aprop_v}
\end{align}
where the propagation of the field corresponds to the sum over all the atomic operators with velocity $v_z$ and Maxwell distribution $f(v_z)$.
Therefore, for the steady state, we obtain the following equation
\begin{align}
\frac{\partial \mathbb{\hat{A}}(z)}{\partial z}&=\mathbb{R}~\mathbb{\hat{A}}(z),\label{eq:dA_dz_dopp}
\end{align}
such that the amplitude of the fields at the output of the propagation is given by
\begin{align}
\mathbb{\hat{A}}(z)&=e^{\mathbb{R}z}\mathbb{\hat{A}}(0),\label{eq:SolA_z_dopp}
\end{align}
with
\begin{align}
\mathbb{R}=\frac{ N\ \mathbb{T}}{c} \int dv_z f(v_z)\mathbb{M}(v_z)^{-1}\mathbb{G}_x
\end{align}

From this expression, we can integrate the output fields with respect to the longitudinal direction of the propagation of the fields and determine, from eq.(\ref{eq:gain_modes_f}), the gain distribution of the modes.

\section{Convergence of the Floquet expansion \label{app:Convergence} }
The Floquet expansion in eq.(\ref{eq:XFloqExp_z}) achieves convergence when a large number of modes are considered. In the ideal case, we would choose a number that is large enough to guarantee convergence. However, this might require extreme computational resources, which are not always at our disposal. Therefore, we can set a finite range of cut-off modes that lead to convergence within a range of tolerance.

Figure~\ref{fig:convergence_ang_3mrads} shows the convergence of the Floquet expansion for the gain distribution as a function of the two-photon detuning, which is the main parameter of the interaction. Here, we  considered the angle $\theta_{\mathrm{eff}}=3$~mrads, which is an interesting case because it is expected to produce a large number of modes. Thus, Fig.~\ref{fig:convergence_ang_3mrads}(a) and (b) show the gain distribution of modes for the probe and conjugate channels, respectively.  The main modes with non-zero gain $(\hat{a}_0,\hat{a}_2,\hat{a}_4,\hat{a}_6)$ and  $(\hat{b}_0,\hat{b}_2,\hat{b}_4,\hat{b}_6)$ are plotted with solid lines, which correspond to the calculation with cut-off modes $Q=20$, whereas the curves with dashed lines correspond to the calculation with $Q=18$. The modes $(\hat{a}_{-2},\hat{a}_{-4},\hat{a}_{-6})$  and $(\hat{b}_{-2},\hat{b}_{-4},\hat{b}_{-6})$ are not shown because of symmetry (they have the same gain as the positive modes) and to simplify the figures. In the sequence, Figs.~\ref{fig:convergence_ang_3mrads}(c-j) show the residuals for each mode and each channel (probe and conjugate) comparing the calculations of $Q=20$ and $Q=18$.
\begin{center}
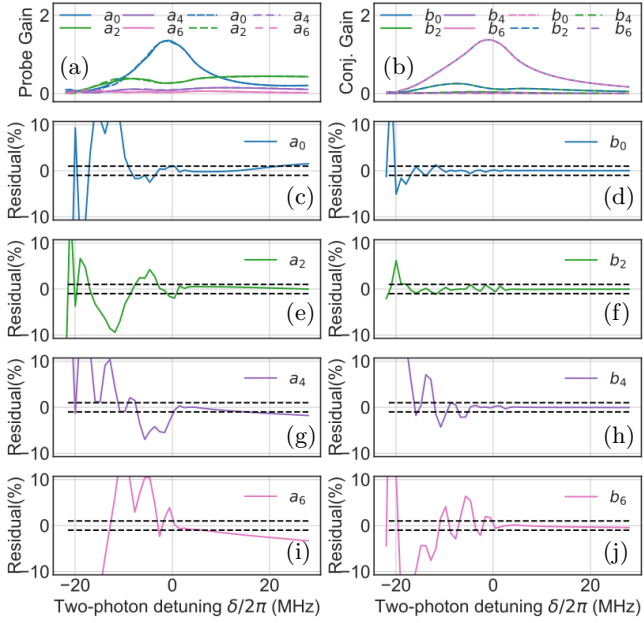
\begin{figure}[h!]
\begin{overpic}[width=0.48\textwidth]{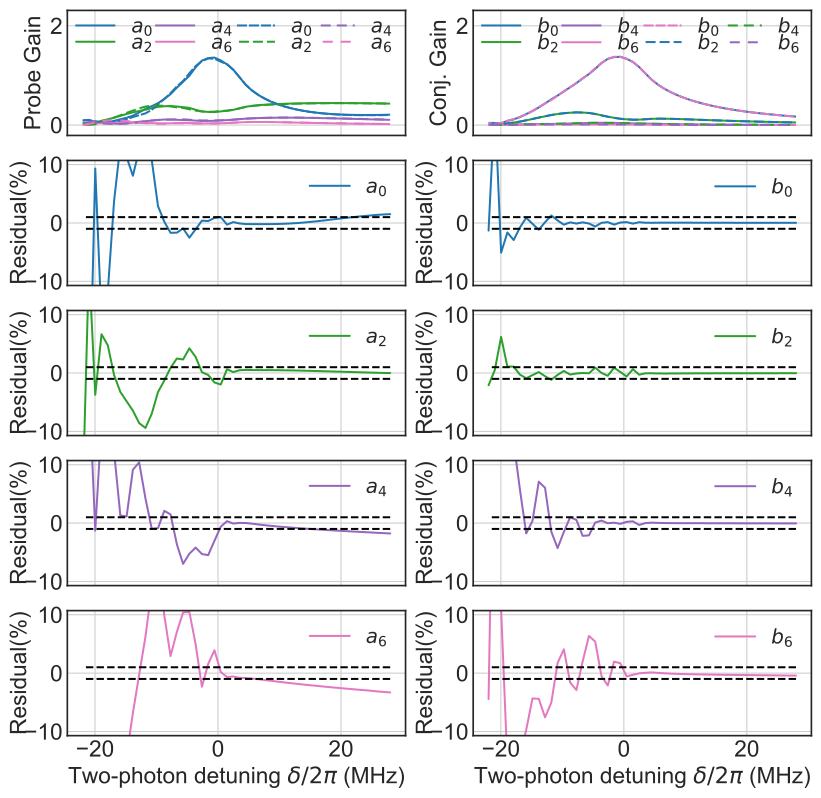}
\put(9,85){(a)}
\put(59,85){(b)}
\put(44,65){(c)}
\put(93,65){(d)}
\put(44,47){(e)}
\put(93,47){(f)}
\put(44,28){(g)}
\put(93,28){(h)}
\put(44,10){(i)}
\put(93,10){(j)}
\end{overpic}
\caption{Gain distribution for the probe and conjugate channel. (a) and (b) for $\theta_{\mathrm{eff}}=3$~mrads, (c) and (d) for $\theta_{\mathrm{eff}}=4.8$~mrads and (e) and (f) for $\theta_{\mathrm{eff}}=6$~mrads. The parameters of the calculation are: $\Omega_0/2\pi=220$~MHz, $\Delta/2\pi=0.9$~GHz, $\Gamma/2\pi=5.7$~MHz,  $\gamma_d/2\pi=1$~MHz, $\omega_{\mathrm{HF}}/2\pi=3.035$~GHz and $g_a=g_b=0.28$~MHz.}
\label{fig:convergence_ang_3mrads}
\end{figure}
\end{center}

From Figs.~\ref{fig:convergence_ang_3mrads}(a) and (b), we can observe that the gain represented by dashed-lines is superposed for a wide range of two-photon detunings. However, a more quantitative variation can be observed with the residuals for each mode. Figures~\ref{fig:convergence_ang_3mrads}(c) and (d)  show the residuals for modes  $(\hat{a}_0$ and $(\hat{b}_0$, respectively. Notice that the residuals are within 1\% for $\delta/2\pi>0$~MHz, with a slight variation up to 1.5\% for large detunings $\delta > 20$~MHz on the probe channel, whereas they remain constant below 1\% for the conjugate channel. In contrast, for $-20<\delta/2\pi<0$~MHz, the residuals fluctuate within a range of 10\% for the probe channel while remaining within 1\% for the conjugate channel. This shows that for the number of mode $Q=20$ the level of confidence on the solution is enough up to 1\% tolerance for $\delta/2\pi>0$ and up to 10\% for $\delta/2\pi<0$. The reason of how it presents higher fluctuation in the convergence for $\delta/2\pi<0$ is that in that region, the probe and the conjugate are considerably attenuated as the seed beam is further red detuned. More specifically, the modes are tuned to the region of Raman transition, where the probe and conjugate modes are absorbed to produce light in the pump field, as in the case of a single pump.

Now, if we look at the residuals for $\hat{a}_{2}$ ($\hat{b}_{2}$), $\hat{a}_{4}$ ($\hat{b}_{4}$) and $\hat{a}_{6}$ ($\hat{b}_{6}$), in Figs.~\ref{fig:convergence_ang_3mrads}(e) to (j), respectively, the variation remains within 1\% for $\delta/2\pi>0$, except for $\hat{a}_{6}$ ($\hat{b}_{6}$), which changes by up to 2\% for $\delta/2\pi>10$~MHz. On the other side of the spectrum, for $-10<\delta/2\pi<0$~MHz the residual presents a variation within a range of 10\%, which means that for some detuning the variation is less, by 2\% or 5\%. For $\delta/2\pi<-10$~MHz, the variation can exceed 10\% because the gain for both channels approaches zero, which would require a larger number of cut-off modes $Q$ to achieve convergence with lower values of residuals.

Since the main properties of quantum correlation and entanglement occur for $\delta/2\pi>0$,  this characterisation indicates that by using $Q=20$ we can obtain a solution with a level of confidence of about 1\%. Since the calculation involves Doppler integration, the mode space with $Q=20$ finds a balance between the confidence on the solution and the computational cost or time consumption for the calculation.

\section{Stochastic noise \label{app:stc_noise}}
The solution for the propagation of the fields in eq.(\ref{eq:d_deltaAFloqExp}) is given by
\begin{align}
\delta\mathbb{\hat{A}}(z,\omega)&=\mathbb{J}(z,\omega)
\delta\mathbb{\hat{A}}(0,\omega)+\mathbb{J}(z,\omega)
\mathbb{F}_{in}(z,\omega)\label{eq:SolFFTdA_n_app}
\end{align}
with the first term as the amplification of the input field and the second term as the amplification of vacuum fluctuations, where
\begin{align}
\mathbb{F}_{in}(z,\omega)=\int_0^z dz' e^{-\mathbb{R}(\omega)\ z'}\mathbb{R}_F(\omega)\mathbb{F}(\omega,z') \label{eq:stc_sol}
\end{align}

From this solution, the power spectral matrix for the amplitude field operators is given by 
\begin{align}
\mathbb{S}(z,\omega)=(2\pi c/L) \Braket{\delta\mathbb{\hat{A}}(z,\omega),\delta\mathbb{\hat{A}}(z,\omega')^T}\delta(\omega+\omega')\label{eq:SolFFTS}
\end{align}
such that, using the solution in eq.(\ref{eq:SolFFTdA_n}) we obtain
\begin{align}
\mathbb{S}(z,\omega)=&\left[ \mathbb{J}(z,\omega)\mathbb{S}(0,\omega)\mathbb{J}(z,\omega')^T\right.\nonumber\\
&+\left.\mathbb{J}(z,\omega)\mathbb{S}_F(z,\omega)\mathbb{J}(z,\omega')^T\right]
\end{align}
where the first term corresponds to the input power spectral density
\begin{align}
\mathbb{S}(0,\omega)&=(2\pi c/L)\Braket{\delta\mathbb{\hat{A}}(0,\omega),\delta\mathbb{\hat{A}}(0,\omega')^T}\delta(\omega+\omega')\nonumber\\
&=2\pi \mathbb{S}_A(\omega) \delta(\omega+\omega')
\end{align}
and the second term corresponds to the power spectral density matrix from the stochastic term, such that
\begin{align}
\mathbb{S}_F(z,\omega)&=(2 c/L)\Braket{\mathbb{\hat{F}}_{in}(z,\omega),\mathbb{\hat{F}}_{in}(z,\omega')^T}
\end{align}

The cross terms between the input modes $\delta\mathbb{\hat{A}}(0,\omega)$ and the stochastic $\mathbb{\hat{F}}_{in}(z,\omega)$ are not correlated and, therefore, do not contribute to the total power spectral density matrix.

From the solution in eq.(\ref{eq:stc_sol}), the spectral density of the stochastic term is given by
\begin{widetext}
\begin{align}
\Braket{\mathbb{\hat{F}}_{in}(z,\omega),\mathbb{\hat{F}}_{in}(z,\omega')^T}&=\int_0^z\int_0^z dz' dz'' e^{-\mathbb{R}(\omega)\ z'}\mathbb{R}_F(\omega) \Braket{\mathbb{\hat{F}}(z',\omega),\mathbb{\hat{F}}(z'',\omega')^T}\mathbb{R}_F(\omega')^Te^{-\mathbb{R}(\omega)^T\ z''}\label{eq:psd_stc}
\end{align}
\end{widetext}
which describes a spectral correlation of the stochastic operator for different modes $(n)$ in the frequency domain. This spectral correlation is calculated from the autocorrelation function of the stochastic operator  in the time domain, which is determined by the diffusion matrix, such that
\begin{align}
\Braket{\mathcal{\hat{F}}_i^{(n)}(z,t),\mathcal{\hat{F}}_j^{(m)}(z',t')}&=\mathbf{D}_{ij}^{(n)}\delta(z-z')\delta(t-t')\delta_{n,m=-n}
\end{align}
which is a second order correlation of the stochastic operator for instantaneous times $t'=t$. The auto correlation function take  non-zero values for modes $m=-n$.

In the frequency domain,  the spectral  correlation is expressed as
\begin{align}
\Braket{\mathcal{\hat{F}}_i^{(n)}(z,\omega),\mathcal{\hat{F}}_j^{(n)}(z',\omega')}&=2\pi\mathbf{D}_{ij}^{(n)}\delta(z-z')\delta(\omega+\omega')
\end{align}
($\mathbf{D}_{ij}^{(n)}=\mathbf{D}_{ij}^{(0)}\delta_{n,-n}$).
Hence, the power spectral matrix from the stochastic operator in eq.(\ref{eq:psd_stc}) can be written as 
\begin{widetext}
\begin{align}
\Braket{\mathbb{\hat{F}}_{in}(z,\omega),\mathbb{\hat{F}}_{in}(z,\omega')^T}&=2\pi\delta(\omega+\omega')\int_0^z dz'  e^{-\mathbb{R}(\omega)\ z'}\mathbb{R}_F(\omega) \mathbb{D}\mathbb{R}_F(\omega')^Te^{-\mathbb{R}(\omega')^T\ z'}=2\pi \tilde{\mathbb{D}}(z,\omega,\omega')\delta(\omega+\omega')
\end{align}
where the multimode difusion matrix is given by
\begin{align}
\tilde{\mathbb{D}}(z,\omega,\omega')=\begin{bmatrix}
   0 & 0 & \cdots & 0 & \cdots & 0 & \tilde{\mathbf{D}}^{(-Q)}(z,\omega,\omega')\\
   0 & 0 & \cdots & 0 & \cdots & \tilde{\mathbf{D}}^{(-Q+1)}(z,\omega,\omega') & 0 \\
    0 & 0 & \cdots & 0 & \cdots & \cdots & 0 \\
     0 & 0 & \cdots & \tilde{\mathbf{D}}^{(0)}(z,\omega,\omega') & \cdots & 0 & 0 \\
      0 & 0 & \cdots & 0 & \cdots & 0 & 0 \\     
  0 & \tilde{\mathbf{D}}^{(Q-1)}(z,\omega,\omega') & \cdots & 0 & \cdots & 0 & 0 \\
     \tilde{\mathbf{D}}^{(Q)}(z,\omega,\omega') & 0 & \cdots &0 & \cdots& 0 & 0 
\end{bmatrix}
\end{align}
with each matrix element
\begin{align}
\tilde{\mathbf{D}}^{(n)}(z,\omega,\omega')&=\begin{bmatrix}
   D_{a,a}^{(n)}(z,\omega,\omega') & D_{a,a^\dagger}^{(n)}(z,\omega,\omega') & D_{a,b}^{(n)}(z,\omega,\omega') & D_{a,b^\dagger}^{(n)}(z,\omega,\omega')\\
    D_{a^\dagger,a}^{(n)}(z,\omega,\omega') & D_{a^\dagger,a^\dagger}^{(n)}(z,\omega,\omega') & D_{a^\dagger,b}^{(n)}(z,\omega,\omega') & D_{a^\dagger,b^\dagger}^{(n)}(z,\omega,\omega')\\
    D_{b,a}^{(n)}(z,\omega,\omega') & D_{b,a^\dagger}^{(n)}(z,\omega,\omega') & D_{b,b}^{(n)}(z,\omega,\omega') & D_{b,b^\dagger}^{(n)}(z,\omega,\omega')\\
     D_{b^\dagger,a}^{(n)}(z,\omega,\omega') & D_{b^\dagger,a^\dagger}^{(n)}(z,\omega,\omega') & D_{b^\dagger,b}^{(n)}(z,\omega,\omega') & D_{b^\dagger,b^\dagger}^{(n)}(z,\omega,\omega')
\end{bmatrix}
\end{align}
\end{widetext}
which is a single pump spectral diffusion matrix. In order to calculate it, we compute the single pump spectral diffusion $\mathbb{D}$ from the Einstein relation with the solution when the atoms interact only with the pump field.

\section{Including losses \label{app:losses}}
We model the presence of losses in the system based on the beam splitter transformation. To do so,  we define $(\mathbb{L})_{n,n}=\text{diag}(\sqrt{\eta},\sqrt{\eta},\sqrt{\eta},\sqrt{\eta})$, with $\eta$ as the detection efficiency that includes losses. Thus, the symmetric covariance matrix follows
\begin{align}
\mathbb{V}^S(z,\omega)'=&\mathbb{L}(\mathbf{V}^S-\mathbb{I})\mathbb{L} + \mathbb{I}
\end{align}
where $\mathbb{I}$ is the covariance matrix of vacuum. If $\eta=1$ then $\mathbb{V}^S(z,\omega)'=\mathbb{V}^S(z,\omega)$, whilst if $\eta=0$ then $\mathbb{V}^S(z,\omega)'=\mathcal{I}$.




\bibliography{refs}






\end{document}